\documentclass[onecolumn,amsmath,amssymb,prl,superscriptaddress]{revtex4}

\usepackage{amsmath}
\usepackage{color}
\usepackage[normalem]{ulem}
\usepackage{natmove}
\usepackage{float}
\usepackage[colorlinks=true, allcolors=blue]{hyperref}
\usepackage{setspace}
\usepackage{graphicx}
\usepackage{multirow}

\begin{document}
	
\title{Exploring the Role of Interdisciplinarity in Physics: Success, Talent and Luck}

\affiliation{Department\ of Physics and Astronomy "Ettore Majorana", University\ of Catania, Italy}
\affiliation{INFN Sezione di Catania, Italy}
\affiliation{Complexity Science Hub Vienna, Austria}
\affiliation{Department\ of Economics and Business, University\ of Catania, Italy}
\affiliation{Department of Clinical and Experimental Medicine, University\ of Catania, Italy}
\affiliation{Institute of Biophysics, Milan, National Research Council of Italy}
\affiliation{Department of Biosciences, University of Milan, Italy}

\author{Alessandro Pluchino}
\affiliation{Department\ of Physics and Astronomy "Ettore Majorana", University\ of Catania, Italy}
\affiliation{INFN Sezione di Catania, Italy}
\author{Giulio Burgio}
\affiliation{Department\ of Physics and Astronomy "Ettore Majorana", University\ of Catania, Italy}
\author{Andrea Rapisarda}
\affiliation{Department\ of Physics and Astronomy "Ettore Majorana", University\ of Catania, Italy}
\affiliation{INFN Sezione di Catania, Italy}
\affiliation{Complexity Science Hub Vienna, Austria}
\author{Alessio Emanuele Biondo}
\affiliation{Department\ of Economics and Business, University\ of Catania, Italy}
\author{Alfredo Pulvirenti}
\affiliation{Department of Clinical and Experimental Medicine, University\ of Catania, Italy}
\author{Alfredo Ferro}
\affiliation{Department of Clinical and Experimental Medicine, University\ of Catania, Italy}
\author{Toni Giorgino}
\affiliation{Institute of Biophysics, Milan, National Research Council of Italy}
\affiliation{Department of Biosciences, University of Milan, Italy}

\author{}
\begin{abstract}

Although interdisciplinarity is often touted as a necessity for modern research, the evidence on the relative impact of sectorial versus to interdisciplinary science is qualitative at best. In this paper we leverage the bibliographic data set of the American Physical Society to quantify the role of interdisciplinarity in physics, and that of talent and luck in achieving success in scientific careers. We analyze a period of 30 years (1980-2009) tagging papers and their authors by means of the Physics and Astronomy Classification Scheme (PACS), to show that some degree of interdisciplinarity is quite helpful to reach success, measured as a proxy of either the number of articles or the citations score. We also propose an agent-based model of the publication-reputation-citation dynamics reproduces the trends observed in the APS data set. On the one hand, the results highlight the crucial role of randomness and serendipity in real scientific research; on the other, they shed light on a counterintuitive effect indicating that the most talented authors are not necessarily the most successful ones. \\ 

\end{abstract}

\maketitle

The importance and the beneficial role of interdisciplinarity is very  often advocated in editorials of high impact journals, public speeches about innovative research policies and research proposal guidelines 
 \cite{Van-noorden,Rylance,Carley,Nas,Saramaki}.  
 However, starting and pursuing interdisciplinarity research projects is fraught with difficulties and risks. First of all, funds are difficult to obtain because the evaluation of projects is frequently underestimated or misjudged, since they pertain to different disciplines, thus requiring a much broader assessment. Secondly, interdisciplinary groups are somehow risky for young researchers, because a hybrid curriculum does not help in getting career advancements. Thirdly, developing a common language among scientists with different backgrounds is tough and very time demanding.

Despite such difficulties, there are several indications that  the interdisciplinary character of research is growing and that this  can be considered a positive signal for the progress of science \cite{Sinatra15,Latora}.
In the last decades we have seen the birth of new disciplines originated by the contamination of physics with biology, computer science and big data, finance, economics, and social sciences in general \cite{Stanley,Helbing,Castellano,Schwe}.  
Most of this hybridization is due to the acknowledgement of the complex nature of socio-economic phenomena \cite{Nature2018}, which in turn fostered research on the field of complex systems in universities and research centers, attracting scientists coming from different backgrounds. In this context, physicists and economists have recently tried to understand the determinants of innovation and success 
\cite{Pietronero,Cimini,Loreto,Barabasi,Sinatra15,Fortunato,Sinatra16,Pluchino18,FRANK,Stanley2} from a new and rigorous perspective. In this study we investigate the role of interdisciplinarity in physics research and question how much it is a key ingredient for a successful career. 
In particular, we address the publication-reputation-citation dynamics of the physicists research community by exploiting the information extracted from the  American Physical Society (APS) data set. In the first part of the paper, we evaluate how much the individual propensity to interdisciplinarity influences the score of an author in terms of both publications and citations. In the second part, inspired by a recent study \cite{Pluchino18}, we present an agent-based model that quantitatively reproduces the stylized facts of empirical APS data. Finally, we try to analyse  the role of chance in reaching high levels of success in scientific careers.


\section*{Role of Interdisciplinarity in the APS data set}

The APS data set is a corpus of articles published in Physical Review Letters, Physical Review and Reviews of Modern Physics, and dates back to 1893 (see Supplementary Information S1 for more details about this section). In particular, data about of all citing-cited pairs of articles in which one paper cites another within the collection, basic metadata and sub-disciplinary classifications about each article in the collection, are present. Data in APS data set were preprocessed and cleaned to avoid duplicated authors and affiliations by using Jaccard similarity in connection to Locality Sensitive Hashing. For the aims of our analysis we will only consider the period $1980-2009$ and the articles of the $N=7303$ authors who published their first paper between $1975$ and $1985$ and at least three papers between $1980$ to $2009$. Their total scientific production in this interval of $30$ years consists of $89949$ PACS classified articles, which received a total of $1329374$ citations from the other articles in the data set. The Physics and Astronomy Classification Scheme (PACS) is a hierarchical partitioning of the whole spectrum of subject matter in physics, astronomy, and related sciences introduced since $1975$. We limit our attention to the 10 most general PACS classes, each corresponding to a broad disciplinary field. Information about classes present in each article are available in the APS data set.    

\begin{figure}[t]
\begin{center}
\includegraphics[width=3.6 in,angle=0]{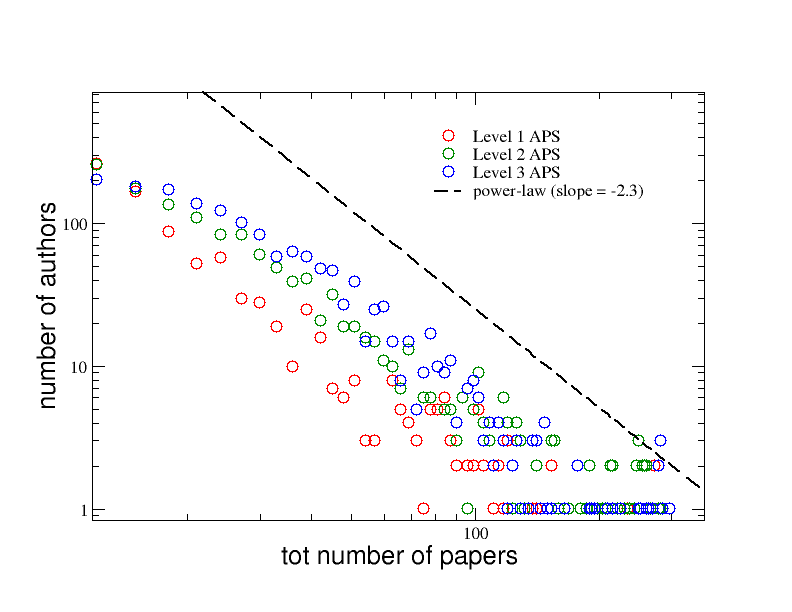}
\caption{\small 
{\it APS data set}. Distributions of the total number of papers published, during their entire careers, by the authors of the three groups with increasing levels of interdisciplinarity. A power-law curve with slope equal to $-2.3$ is also reported for comparison (dashed line).       
}
\label{papers-real} 
\end{center}
\end{figure}

For a given author $A_i$ ($i=1,...,N$), the total number $D^{APS}_i \in [1,10]$ of different PACS classes appearing in her publications during her entire career could be certainly considered as a global indicator of the multidisciplinarity of her work. However, as the $D^{APS}_i$ classes do not appear simultaneously in all the papers of $A_i$, it is also interesting to consider the average number $d^{APS}_i \in \mathbb{R}$ of classes simultaneously present in her articles. From the APS data it results that $d^{APS}_i \in [1,3.33]$. Multiplying these two factors, we finally obtain the index 
\begin{equation}
I^{APS}_i = D^{APS}_i \times d^{APS}_i	
\end{equation}
which we propose as a more robust indicator of interdisciplinarity.

Accordingly, we can divide the $N=7303$ authors into three groups, with comparable sizes and with an increasing level of interdisciplinarity: 

- Level 1 group $L_1^{APS}$: $N_1=2445$ authors with $1 \le I^{APS}_i \le 3$ (low interdisciplinarity level);

- Level 2 group $L_2^{APS}$: $N_2=2511$ authors with $3 < I^{APS}_i \le 6$ (medium interdisciplinarity level);

- Level 3 group $L_3^{APS}$: $N_3=2347$ authors with $I^{APS}_i > 6 $ (high interdisciplinarity level);

The first goal of this study is to investigate if these different degrees of interdisciplinarity are correlated to the scientific impact of the active researchers of the APS data set, evaluated through both the number of papers and the citations cumulated during their careers. 

\begin{figure}[t]
\begin{center}
\includegraphics[width=3.6 in,angle=0]{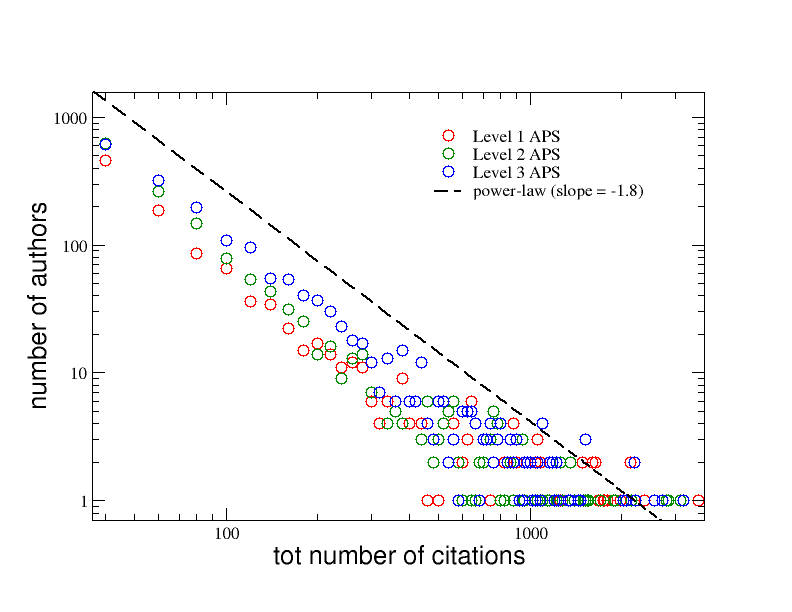}
\caption{\small 
{\it APS data set}. Distributions  of the total number of citations cumulated, during their entire careers, by the authors of the three groups with increasing levels of interdisciplinarity. A power-law curve with slope -1.8 is also reported for comparison (dashed line).       
}
\label{citations-real} 
\end{center}
\end{figure}

Fig.\ref{papers-real} shows the distributions of the total number of papers published by the authors belonging to the three groups $L_1^{APS}$, $L_2^{APS}$ and $L_3^{APS}$, which are plotted in three different colors (red, green and blue, respectively). The interdisciplinarity level has a strong positive influence on the productivity of the authors, since authors that have a higher level of  interdisciplinarity are more productive. In addition, the tails of  the three distributions follow a power-law behavior, with a slope equal to $-2.3$ (dashed line). A similar  behavior is visible in Fig.\ref{citations-real}, where the distributions of the total number of citations received by the authors of the three groups during their entire careers is plotted with the same colors. Also in this case, the interdisciplinarity level seems to play an important role in affecting the scientific success of the researchers. Again, the tails of the three distributions follow a power-law behavior, but here with a different slope equal to $-1.8$.     

\begin{table}[]
\begin{tabular}{c|c|c|c|c|c|}
\cline{2-6}
& \,Authors\, & \,Papers\, & \, $\bar{P}^{APS}$ \, & \,Citations\, & \, $\bar{C}^{APS}$ \, \\ \hline
\multicolumn{1}{|c|}{$L_1^{APS}$} & 2445 & 18832  & 7.70 & 230448 & 94.25\\ \hline
\multicolumn{1}{|c|}{$L_2^{APS}$} & 2511 & 35892  & 14.29 & 515635 & 205.35\\ \hline
\multicolumn{1}{|c|}{$L_3^{APS}$} & 2347  & 50947  & 21.71 & 843292 & 359.31\\ \hline
\end{tabular}
\caption[]{
				\label{real_details}
				 Some characteristic numbers of our APS sample regarding  the 89949 published papers and  their 1329374 citations over the three interdisciplinarity groups. Here $\bar{P}^{APS}$ = average papers per author and  $\bar{C}^{APS}$ = average citations per author. A paper is counted in more than one class if it is coauthored by researchers belonging to different classes, so the sums of the number of papers and of the citations exceed, respectively, 89949 and 1329374.}
\end{table}

\begin{figure}[H]
\begin{center}
\includegraphics[width=3.57 in,angle=0]{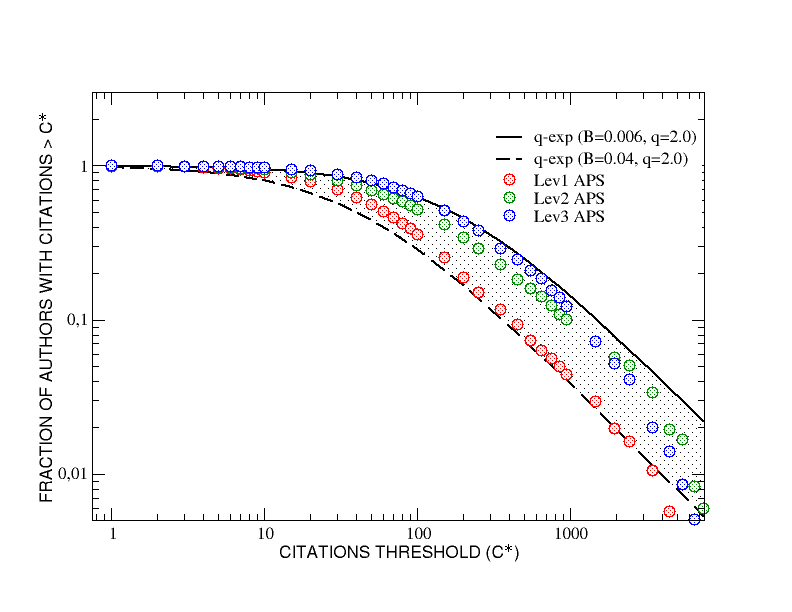}
\caption{\small 
{\it APS data set}. The fraction of researchers who have collected, during their careers, a number of citations greater than an increasing threshold $C^*$ is reported as function of  $C^*$. The three curves, corresponding to different interdisciplinarity levels, lie inside a range limited by two q-exponential functions with the same value for the entropic index $q$ and different values of $B$ as reported.
}
\label{q-citations} 
\end{center}
\end{figure}

In Table \ref{real_details} we report the total number of papers and citations for each interdisciplinarity group, together with the corresponding averages per author ($\bar{P}$ and $\bar{C}$, respectively). The results confirm the hypothesis that $I^{APS}_i$ is able to capture a real interesting effect encoded in the APS data set, i.e. the beneficial role of interdisciplinarity in enhancing both the productivity and the scientific impact of the examined authors.        

In order to better appreciate the differences between the three interdisciplinarity groups from the point of view of their citations scores, it is convenient to take into account, for each group, the fraction of authors who have collected, during their whole careers, a number of citations greater than an increasing threshold $C^*$. In Fig.\ref{q-citations} such a cumulative citation distribution is plotted as function of the threshold $C^*$, for the three interdisciplinarity groups. The fraction of highly interdisciplinary researchers (level 3) is generally greater that the fraction of those of level 2 and level 1, in particular for threshold values less than $C^*=1000$. Above this value, a tendency towards a mixing of the three groups is visible; in any case, authors of level 3 stay always above those of level 1. It is also interesting to notice that the three curves lie inside a range limited by two q-exponential functions $y = [1-(1-q)Bx]^{\frac{1}{1-q}}$ \cite{Tsallis} with the same value for the entropic index $q=2.0$ and different values of $B$, see Fig.\ref{q-citations}.

\begin{figure}[t]
\begin{center}
\includegraphics[width=3.0 in,angle=0]{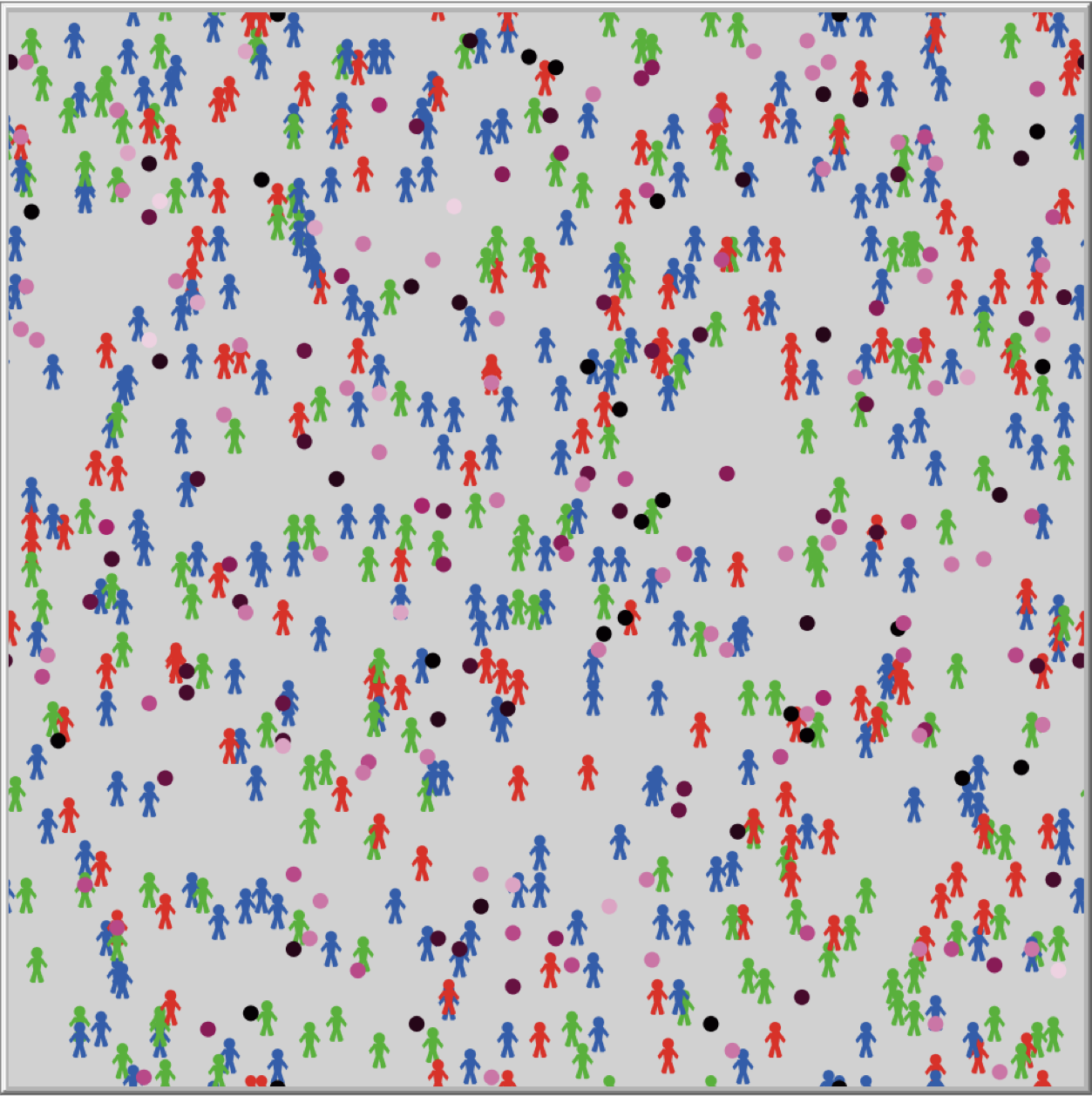}
\caption{\small 
A simplified depiction of the initial state of the agent-based model. For clarity, the figure represents only $500$ individuals but the simulations considered the whole cohort of $N=7303$ researchers active in the period of $30$ years taken into account in the analysis of the APS data set. The 'world' is a 2D box with periodic boundary conditions.
}
\label{world-main} 
\end{center}
\end{figure}

\section*{The Agent-Based Model}

Next, we turned to the development of an agent-based model able to reproduce, under constraints based on APS real data, the publication-reputation-citation dynamics which generates the observed behavior of our cohort of scholars (see Supplementary Information S2 for more details about this section). 

We start by choosing the initial setup of the model in order to take into account some of the real features of the authors considered in the APS data set. In Fig.\ref{world-main} we show the 2D model world where the $N=7303$ APS authors (agents depicted as silhouettes) are randomly assigned a position, fixed during the simulations. Each simulation has a duration $t_{max}$ of $30$ years, with a time step $t$ of $1$ year. The agents are divided into the three groups $L_1^{APS}$ (in red), $L_2^{APS}$ (in green) and $L_3^{APS}$ (in blue), with sizes $N_1$, $N_2$ and $N_3$ respectively, according with the value of their real interdisciplinary index $I^{APS}_i = D^{APS}_i \times d^{APS}_i$. Each agent is further characterized by the following other variables: 

- a fixed talent $T_i \in [0,1]$ (intelligence, skill, endurance, hard-working, ...), which is a real number randomly extracted at the beginning of the simulation from a Gaussian  distribution with mean $m_T=0.6$ and standard deviation $\sigma_T=0.1$; 

- an array $\vec{P}_i(t)$, whose elements are the papers published by the author $A_i$ at time $t$; the size $P_i(t)$ of the array $\vec{P}_i(t)$ will give the total number of published papers at that time;  

- an array $\vec{C}_i(t)$, whose elements are the citations received by each paper present in $\vec{P}_i(t)$ at time $t$ (thus, $\vec{C}_i(t)$ and $\vec{P}_i(t)$ have the same size); $C_i(t)$ will be the total number of citations received by all these papers at that time;

- an array $\vec{R}_i(t)$ whose $10$ elements are the reputation levels reached by the author $A_i$ in each one of the PACS classes; these reputation levels are real numbers, included in the interval $[0,1]$, which increase as function of the number of papers published in the corresponding disciplinary fields.   

The virtual world also contains $N_E=2000$ event-points which, unlike the authors, randomly move around during the simulations. They are colored with different shades of magenta, one for each of the $10$ PACS classes, and the relative abundance of points belonging to a given class is fixed in agreement with the information of the APS data set (also their total number $N_E$ was calibrated on the real data).  Events represent random opportunities, ideas, encounters, intuitions, serendipitous events, etc., which can periodically occur to a given individual along her career, triggering a research line along one or more fields represented by the corresponding PACS class. In this respect, each author owns a "sensitivity circle" representing the spatial extension of their sensitivity to the event-points. The radius of these circles is different for the three interdisciplinarity groups $L_1^{APS}$,$L_2^{APS}$, $L_3^{APS}$ and can be determined through a calibration with real data. Each author, depending on her own group, is sensitive only to the event-points corresponding to the "fertile" PACS which are  present in her array $\vec{D}^{APS}_i$ (we will define these points as "special" points for that author).  

The publication-reputation-citation dynamics of this community of authors is quite simple.  
	
- Every year, a check is performed over all the authors in order to verify what type and how many special event-points fall inside their "sensitivity circles". If, for a given author $A_k$ at  year $t$, it results that $0 \le D_k(t) \le D^{APS}_k$ special points are  in her circle, the researcher's talent $T_k$ is compared with a random real number $r \in [0,1]$. 

- If $r<T_k$ (i.e. with probability equal to her talent) the number $P_k(t)$ of her published papers becomes equal to 
\begin{equation}
P_k (t) = P_k (t-1) + \Delta P_k   
\end{equation} 
where $\Delta P_k$ is an integer quantity, randomly extracted from a Gaussian  distribution with mean $m_{P_k} = \mu P_k (t-1)$ and standard deviation $\sigma_{P_k} = \gamma P_k (t-1)$. The increment $\Delta P_k$ is thus proportional to the number of papers already published by $A_k$ during  the previous year $t-1$ (the coefficients $\mu$ and $\gamma$, equal for all the agents, are  fixed through a calibration with the APS data).

- All the newly published papers will be added to the array $\vec{P}_k(t)$ and each of them will be characterized by the PACS classes corresponding to the event-points that fell in the sensitivity circle at year $t$. Thanks to these new publications, author $A_k$ also increases her reputation in each of the disciplinary fields corresponding to their PACS (i.e. the corresponding elements of the array $\vec{R}_k(t)$ will be updated).

- Finally, based on the total number $P_k(t)$ of published papers at time $t$ and on her reputation reached at that time, author $A_k$ yearly updates the elements $c_j(t)$ ($j=1,...,P_k(t)$) of her citations array $\vec{C}_k(t)$ with the following rule: 
\begin{equation}
c_j(t+1) = c_j(t) (1 + \bar{R}_j)    
\end{equation}
In other words, her $j$-th paper will gain citations, at time $t+1$, depending on both its previous citation score $c_j(t)$ and the average reputation $\bar{R}_j$ of the author in the disciplinary fields corresponding to the PACS present in the paper. Therefore, the overall increase in citations for the author $A_k$ at time $t+1$ will be     
\begin{equation}
C_k(t+1) = C_k(t) + \sum_{j=1}^{P_k(t)} c_j(t+1)  
\end{equation}

At the end of the simulation, i.e. for $t=t_{max}$, a generic author $A_k$ will have cumulated a certain number $P_k(t_{max})$ of papers and a certain number $C_k(t_{max})$ of citations, depending on her ability in exploiting the opportunities offered by the random occurrence of event-points within her sensitivity circle. Since this ability is parameterized by the talent, the final success of the researchers - in terms of published papers and cumulated citations - will be influenced by both talent and luck (serendipitous  events). 

\begin{figure}[t]
\begin{center}
\includegraphics[width=3.6 in,angle=0]{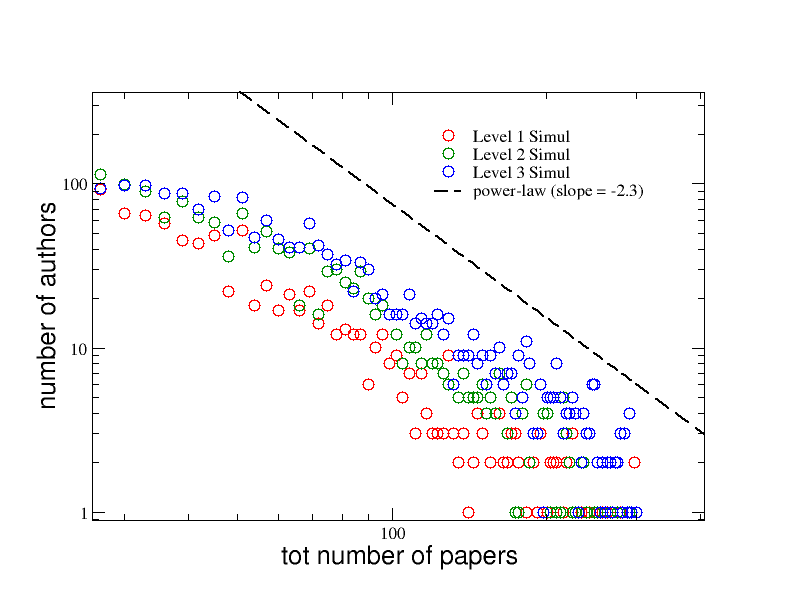}
\caption{\small 
{\it Model Simulation}. Distributions of the total number of papers published, during their simulated careers, by the authors of the three groups with increasing levels of interdisciplinarity. A power-law curve with slope -2.3 is also reported for comparison (dashed line).       
}
\label{papers-simul} 
\end{center}
\end{figure}

\section*{Numerical results}

In this section we are interested in verifying if our model is able to capture the stylized facts already observed in the APS data set, with particular regard to the role of interdisciplinarity. 
Before going on, it is important to note that the agent-based model allows one to
 average the number of PACS present in each of the $P_i(t_{max})$ publications of a given author $A_i$ at the end of a simulation, enabling the calculation of the dynamical counterpart of the real parameter $d^{APS}_i$, i.e. the new parameter $d^{sim}_i(t_{max})$. By multiplying this parameter for the real $D^{APS}_i$, it is possible to update the interdisciplinarity index $I^{APS}_i$ of Eq.1 - assigned at the beginning of the simulation on the basis of the real APS data - therefore obtaining the new (simulated) index
\begin{equation}
	I^{sim}_i = D^{APS}_i \times d^{sim}_i
\end{equation}
This index, which quantifies the effective interdisciplinarity level reached by each author at the end of a simulation, will allow, in turn, to update also the membership of the authors to the three interdisciplinarity groups, which now become $L_1^{sim}$, $L_2^{sim}$ or $L_3^{sim}$. 

As function of these groups, i.e. of the three corresponding interdisciplinarity levels, we plot in Fig.\ref{papers-simul} the distributions of the total number $P_i(t_{max})$ of papers published by the $N$ authors at the end of a typical simulation. We adopt the same colors as in Fig.\ref{papers-real}: red, green and blue, respectively. It is evident from the plot, that the proposed model is able to reproduce the same kind of behavior observed for the real APS data set: again, the degree of interdisciplinarity seem to have a strong positive correlation with  the productivity of the authors and also the tails of the three distributions follow a power-law trend with the same slope of $-2.3$ found for the APS data set. 

\begin{figure}
\begin{center}
\includegraphics[width=3.6 in,angle=0]{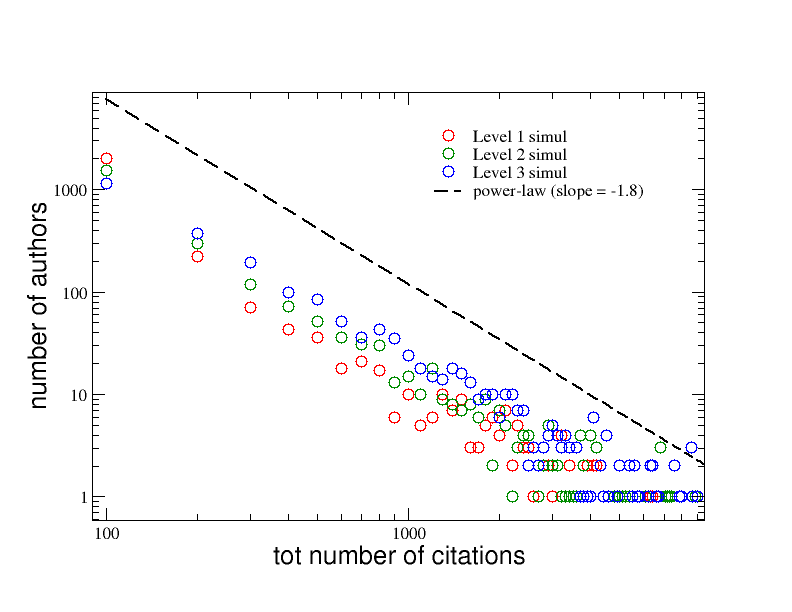}
\caption{\small 
{\it Model Simulation}. Distributions of the total number of citations cumulated, during their simulated careers, by the authors of the three groups with increasing levels of interdisciplinarity. A power-law curve with slope -1.8 is also reported for comparison (dashed line).       
}
\label{citations-simul-main} 
\end{center}
\end{figure}

An analogous agreement with the APS data can be observed in Fig.\ref{citations-simul-main}, where we show the distributions of the total number $C_i(t_{max})$ of citations cumulated by the authors of the three groups during the simulation of their careers. Also in this case, the scientific impact seems strictly correlated with the interdisciplinarity propensity of the researchers. Moreover, the tails of the three distributions follow the same power-law behavior observed for the APS  data (see Fig.\ref{citations-real}), with a slope of $-1.8$. 

\begin{figure}
\begin{center}
\includegraphics[width=3.5 in,angle=0]{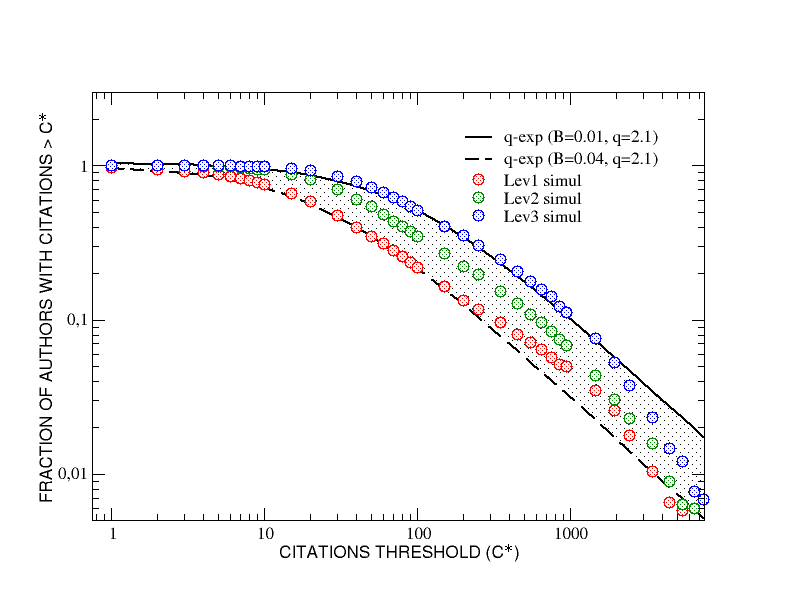}
\caption{\small 
{\it Model Simulation}. The fraction of researchers who have collected, during their simulated careers, a number of citations greater than an increasing threshold $C^*$ is reported as function of  $C^*$. The three curves, corresponding to different interdisciplinarity classes, lie inside a range limited by two q-exponential functions with the same value for the entropic index $q$ and different values of $B$. 
}
\label{q-citations-simul} 
\end{center}
\end{figure}

\begin{figure}[t]
\begin{center}
\includegraphics[width=3.6in,angle=0]{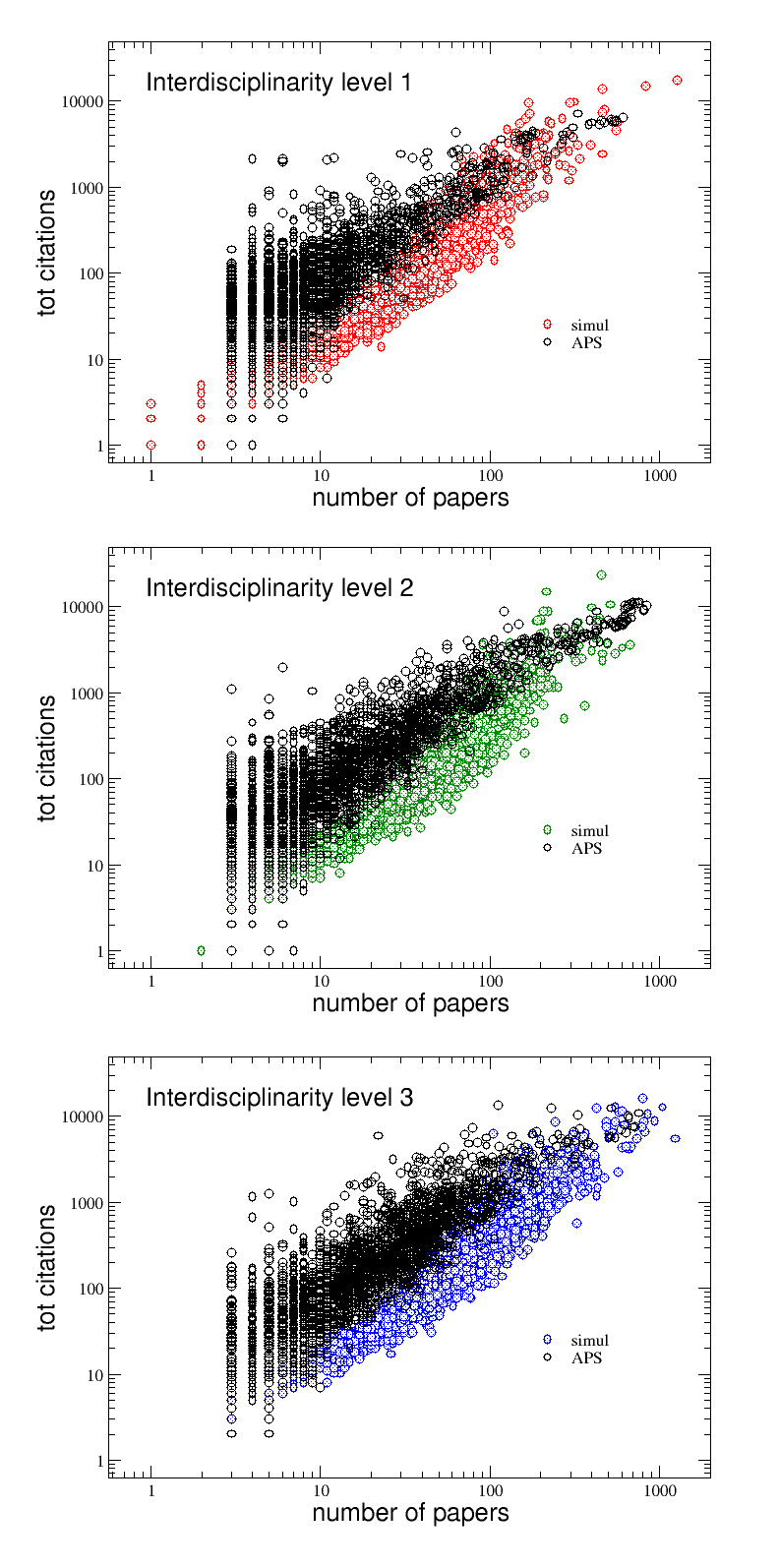}
\caption{\small 
The total number of citations of the researchers of the three interdisciplinarity levels as function of the total number of papers they published in their careers. The figures show that the agent-based model numerical simulations are able to reproduce the positive correlation between these two quantities in all the three groups.
}
\label{citations-vs-papers} 
\end{center}
\end{figure}

It is also interesting to plot, as in the previous section, the fraction of authors of the three interdisciplinarity groups who have collected, during their whole simulated careers, a number of citations greater than an increasing threshold $C^*$. In Fig.\ref{q-citations-simul} it results that the fraction of highly interdisciplinary researchers (level 3) is greater than the fraction of those of level 2 for all values of $C^*$, and the latter is - in turn -  always greater than the fraction of level 1 authors. A mixing of the three groups is visible only for very high values of the citation score. As observed in Fig.\ref{q-citations}, also in this case the three curves fall inside a range limited by two q-exponential functions with a value of the entropic index, $q=2.1$, very similar to that obtained for the APS data. 

Finally, in the panels of Fig.\ref{citations-vs-papers}, the positive correlation between $P_i(t_{max})$ and $C_i(t_{max})$ for the APS data set and the agent-based model simulation is shown for the three interdisciplinarity levels.  The agreement between real and simulated data is remarkable. The only feature, visible in the APS data, which the model does not reproduce, is the presence of authors who published just a few papers (even below 10) but with a very high number of citations. Such an occurrence, characterizing in particular the interdisciplinarity level 1, is probably unpredictable since, as it has been shown in a recent study \cite{Sinatra16}, scientists have the same chance of publishing their biggest hit at any moment in their career and even less productive authors have a chance of publishing  very cited papers.

Summarizing, the simulations performed with our agent-based model correctly reproduces the main stylized facts observed in the analysis of the APS data set, confirming that the level of interdisciplinarity plays an important role in determining the scientific success of an author during her academic career. It is important to stress that the calibration of the model with the real data make the output of a single simulation run very robust, despite the differences due to the random initial setup of several model parameters (such as the initial collocation of the authors around the world, their talent distribution and the position/movement of the PACS event-points). This means that these results can be considered quite general and well representative of the model's behavior (we checked that they do not change even performing ensemble averages over many runs).      

Having established the agreement between experimental and modeled data, we turn to the analysis of variables which are impossible to observe directly from the real data. For example, one could wonder if the most successful authors in the three interdisciplinarity groups are also the most talented ones. In a recent numerical study about the causes behind the achievement of success in our life \cite{Pluchino18}, it has been shown that individual talent is necessary but not sufficient to become rich or to climb the social ladder: luck plays a fundamental role and very often moderately gifted, but very lucky, people surpass highly talented, yet unlucky, individuals. 

\begin{figure}[t]
\begin{center}
\includegraphics[width=5.0in,angle=0]{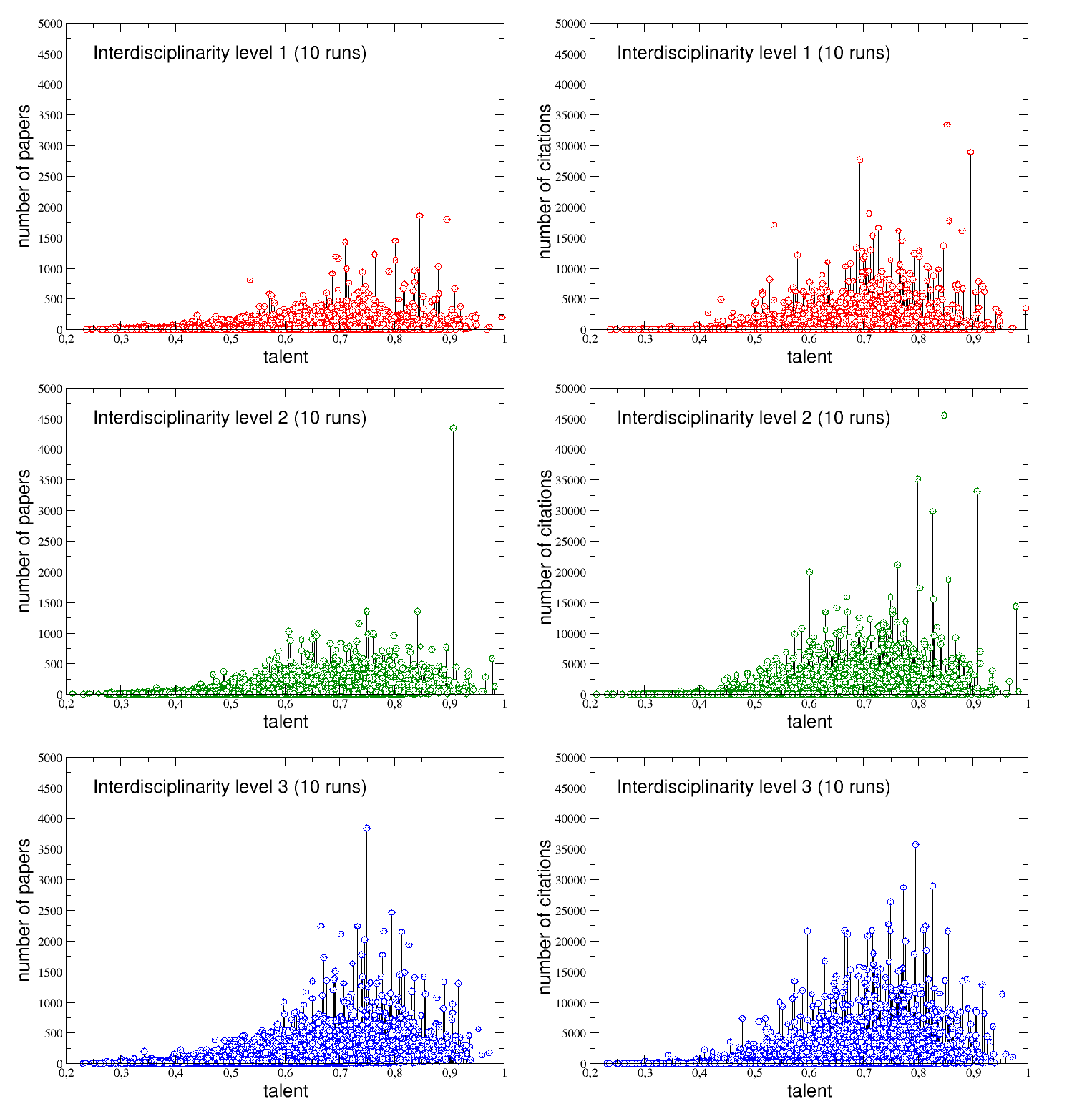}
\caption{\small 
Papers and citations vs talent, collected over 10 replica runs of the same numerical simulation. Each circle in the figures represents the total number of papers (left column) or the total number of citations (right column) cumulated by each author of the three interdisciplinarity groups in each of the 10 runs, reported as function of the corresponding talent. These plots indicate in a clear way that the most successful individuals are never the most talented ones.
}
\label{citations-papers-talent} 
\end{center}
\end{figure}

Finally, we show that this counterintuitive feature holds also in the scientific context addressed here. We performed $10$ replica runs of our agent-based model, with the same calibration based on real APS data, but with different distributions of the talent among the $7303$ authors and with different initial positions for both the agents and the $2000$ event-points. In Fig.\ref{citations-papers-talent} we plot the final number of papers (left column) and the final number of citations (right column) cumulated by each author belonging to the three interdisciplinarity groups during all the $10$ simulations, as a function of their talent. The results indicate that very talented people -- for example researchers with a talent $T_i>0.9$ -- are very rarely the most successful ones, regardless the interdisciplinarity group they belong. Rather, their papers or citations score stays often quite low. On the other hand, scientists with a talent just above the mean -- for example in the range $0.6 < T_i < 0.8$ -- usually cumulate a considerable number of papers and  citations. In other words, the most successful authors are almost always scientists with a medium-high level of talent, rather than the most talented ones. This happens because (i) talent needs lucky opportunities (chances, random meetings, serendipity) to exploits its potentialities, and (ii) very talented scientists are much less numerous than moderately talented ones (being the talent normally distributed in the population). Therefore, it is much easier to find a moderately gifted {\it and} lucky researcher than a very talented {\it and} lucky one.    

\begin{table}
\begin{tabular}{c|c|c|c|c|c|c|c|c|c|}
\cline{2-10}
& \,$\bar{P}^{sim}$\, & \,$N_M$\, & \,$N_T$\, & \,$r_P$\, & \,& \,$\bar{C}^{sim}$\, & \,$N_M$\, & \,$N_T$\, & \,$r_C$\, \\ \hline
\multicolumn{1}{|c|}{$L_1^{sim}$} 
& 30 & $18.6\%$  & $1.1\%$ & 0.06 & & 191 & $8.7\%$ & $0.7\%$  & 0.08 \\ \hline
\multicolumn{1}{|c|}{$L_2^{sim}$} 
& 49 & $18.5\%$  & $1.5\%$ & 0.08 & & 297 & $10.3\%$ & $1.16\%$  & 0.11 \\ \hline
\multicolumn{1}{|c|}{$L_3^{sim}$} 
& 82  & $18.8\%$  & $1.8\%$ & 0.10 & & 505 & $12.1\%$ & $1.5\%$  & 0.12 \\ \hline
\end{tabular}
\caption[]{
				\label{simul_details}
				Details about the percentage of moderately gifted ($N_M$) and highly talented ($N_T$) authors whose publications or citations overcome the respective averages, for each of the three groups with increasing interdisciplinarity level.}
\end{table}

It is also interesting to note that this effect is more pronounced for authors with a low interdisciplinarity level and progressively decreases by increasing the degree of interdisciplinarity. In order to quantitatively address this last point, let us define as moderately gifted ($N_M$) authors $A_i$ with a talent around the mean, i.e. with $0.5<T_i<0.7$, and highly talented ($N_T$) those with $T_i>0.8$ (i.e. greater  than two standard deviations with respect to the mean). Let us also call $\bar{P}^{sim}$ and $\bar{C}^{sim}$ the average values of, respectively, the final number of papers $P_i(t_{max})$ and the final number of citations $C_i(t_{max})$ cumulated by each author inside the three groups $L_1^{sim}$, $L_2^{sim}$, $L_3^{sim}$. Looking to the details in Table \ref{simul_details}, it is evident that, inside each of the three groups, these averages (columns 1 and 5) do increase with the level, highlighting a positive correlation between scientific success and  interdisciplinarity analogous to that one already observed for the same quantities calculated for the APS data set and reported in Table \ref{real_details} (columns 3 and 5). On the other hand, the percentages of highly talented scientists with a final number of papers $P_i(t_{max})>\bar{P}^{sim}$ or with a final number of citations $C_i(t_{max})>\bar{C}^{sim}$, with respect to the same percentages for the moderately gifted one, also increase by increasing the interdisciplinarity level. This is seen by the ratios between the two percentages, $r_P = (N_T / N_M)_P $ and $r_C = (N_T / N_M)_C $, which increase respectively from $0.06$ to $0.1$ and from $0.08$ to $0.12$ going from $L_1^{sim}$ to $L_3^{sim}$. 

\section*{Conclusions}

In conclusion, in this paper we have shown, through both a statistical analysis performed on the APS data set and a comparison with the numerical results obtained by an agent-based model (calibrated on the real data), that the attitude to broaden the scope of their researches, mixing different fields, is able to provide more rewards to the scientists, since their productivity and their scientific impact increase with their level of interdisciplinarity. Moreover, averaging over several runs with different initial distributions of talent among all the authors, we have also shown that, very often, moderately gifted researchers reach higher level of scientific success than very talented ones, simply because they have had more opportunities or just because they were luckier. However, the interdisciplinarity level seems to slightly dampen this effect since its increase does enhance the probability of success of highly talented individuals with respect to the moderately talented ones. Due to the generality of the APS data set, we expect that our findings remain valid beyond the considered case study and beyond physics itself.      

\section*{Acknowledgements}
A.P. and A. R. acknowledge financial support by  the project "Linea di intervento 2" of the  Department of Physics and Astronomy {\it Ettore Majorana} of the University of Catania

\pagebreak

\section{Supplementary Information}

\subsection{S1. APS Data Set Analisys}
	
We give here additional details about the methodologies behind the mining of the American Physical Society (APS) data set from which we have got the results described in the Main Paper (MP). 

The APS data set consists of all the publications of American Physical Society from 1893 to 2013. Each publication is represented through a JSON file storing information about authors, their affiliations, the journal and the PACS or keywords associated to the paper. The database has of more than 550000 publications. A critical aspect relative to the APS data set relies on its noise due to the lexical heterogeneity. Lexical heterogeneity occurs when the tuples have identically structured fields across databases, but the data use different representations to refer to the same real-world object. In our case, authors and affiliations are stored using different conventions in each JSON file. Therefore, the same author, or affiliation can be represented in a different format (i.e. Mark John Smith or Mark J. Smith or Smith M.). Based on this consideration, two records can be considered {\em equivalent} if they are semantically equal. The similarity between records is computed by metrics which measure the semantic equivalence through a score. Record pairs with high similarity scores (above a specified threshold) are treated as duplicates. 

In addition to the accuracy of classifying records pairs into matches and mismatches, the central issue consists of improving the speed of comparisons.
Indeed, cleaning such data before its usage is a mandatory step to avoid redundant and noisy information and affect the reliability of further analysis. To remove duplicate entries we decided to compare two strings (i.e. affiliations of authors) using $q$-grams \cite{qgrams} in connection to Jaccard Similiarity \cite{tan2013data}.  The Jaccard Similarity of two sets $a$ and $b$ is defined as $sim(a,b) = \frac{|a \cap b|}{|a \cup b|}$ ranging from 0 to 1.
Practically, we extracted from each string q-grams of length 2 ($q =2$, for both authors and affiliations), then we claim two authors to be the same when their Jaccard Similarity is greater than a threshold set equal to 0.6. Similarly two affiliations have been declared to be the same if their similarity is greater than the threshold 0.66. These two threshold have been empirically established on a sample of data from APS data set by minimizing the ratio of false negatives (same author/affiliation but we consider the two authors/affiliations as different) and false positives (different authors/affiliations but considered the same author). 
 
Due to the large number of authors and affiliation we experienced a computational bottleneck due to the quadratic time needed to perform all possible pairwise comparisons. To make such a cleaning step feasible we implemented the similarity computation in connection to the Locality Sensitive Hashing (LSH) \cite{cohen2001finding}. LSH is an algorithmic methodology which makes use of hashing, that is able to fast identify similar pairs of objects without comparing them directly. Using such a technique we were able to reduce the computational effort from quadratic to linear.
All the code have been developed in Php and the data, once cleaned, were stored into the relational database MySQL (v. 5.1). Further manipulation and analysis of cleaned data were done using R language.
	
	The measure of the level of interdisciplinarity of the authors (in the discussion we will refer to them also as 'researchers') is based on the APS's PACS ('Physics and Astronomy Classification Scheme'). This scheme consists of a hierarchic partition of the publications in research areas of physics. Any PACS code has four hierarchic levels of increasing specificity: a first and a second digit composing a two-digit number, another two-digit number and a string of characters (e.g. 14.70.Bh). In particular, we work with the less specific hierarchic level, made up by the ten areas of research each corresponding to one of the ten different first digits (0, 1, \ldots 9; or equivalently 00, 10, \ldots 90) of the first two-digit number in the PACS code:

	00 - GP : General Physics

	10 - EPF : Physics of Elementary Particles and Fields
	
	20 - NP : Nuclear Physics
	
	30 - AMP : Atomic and Molecular Physics
	
	40 - EOAHCF : Electromagnetism, Optics, Acoustics, Heat Transfer, Classical Mechanics, and Fluid Dynamics
	
	50 - GPE : Physics of Gases, Plasmas, and Electric Discharges
	
	60 - CM:SMT : Condensed Matter: Structural, Mechanical and Thermal Properties
	
	70 - CM:EEMO : Condensed Matter: Electronic Structure, Electrical, Magnetic, and Optical Properties
	
	80 - IPR : Interdisciplinary Physics and Related Areas of Science and Technology
	
	90 - GAA : Geophysics, Astronomy, and Astrophysics

	Since the APS database regards only the physics' domain, this choice is led by our purpose of identifying an actual interdisciplinarity attitude in the researchers' production. Any published paper can have one or more PACS codes assigned to it and according to our choice we assign different PACS codes to a paper only if these codes differ on the first digit; otherwise, we pile them up on a single code. In this way we assign to each paper a number of PACS codes that is equal to the number of the different broad - less specific - areas related to it. From what has been said, is understood that only PACS classified papers are considered.

	\begin{figure}
		\begin{center}
			\includegraphics[width=.55\linewidth]
		{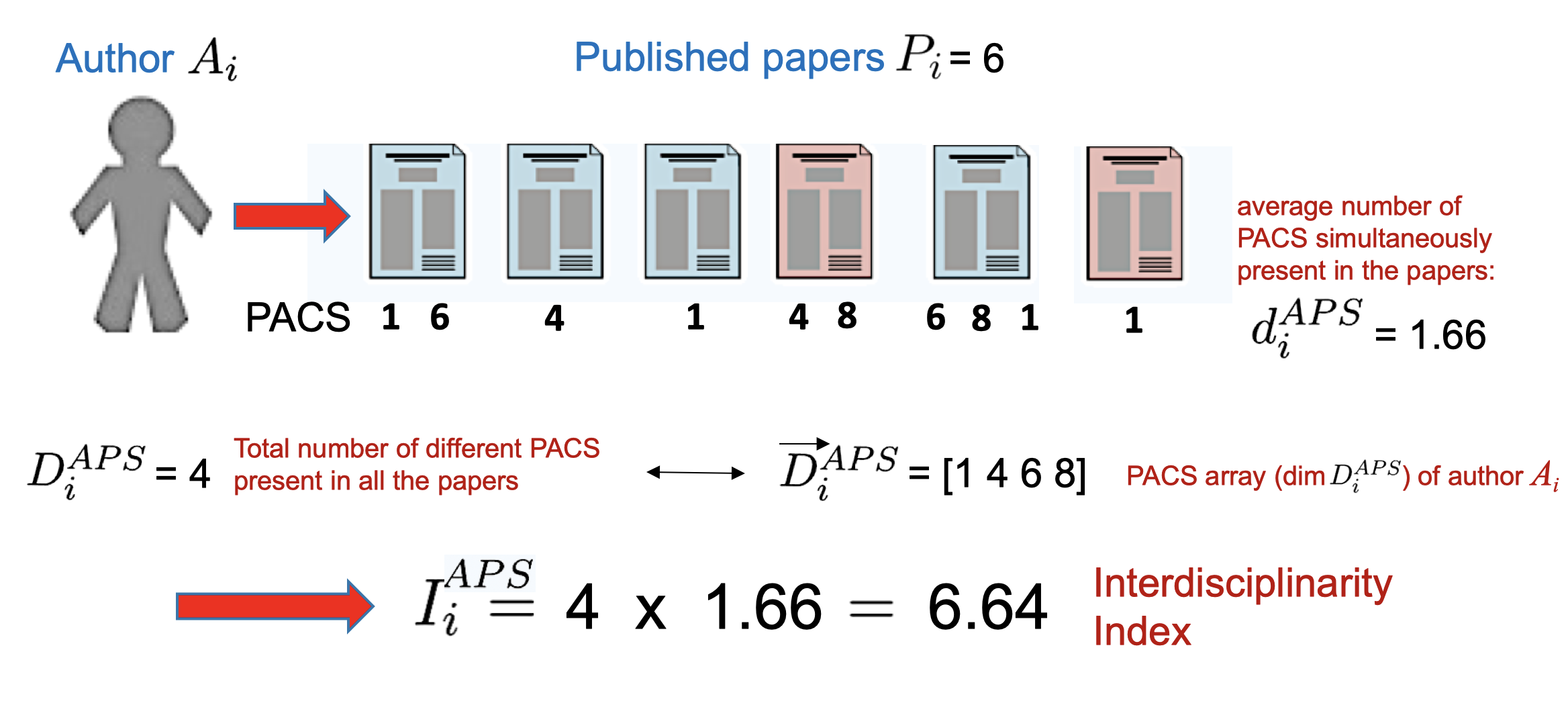}
			\includegraphics[width=.43\linewidth]
		{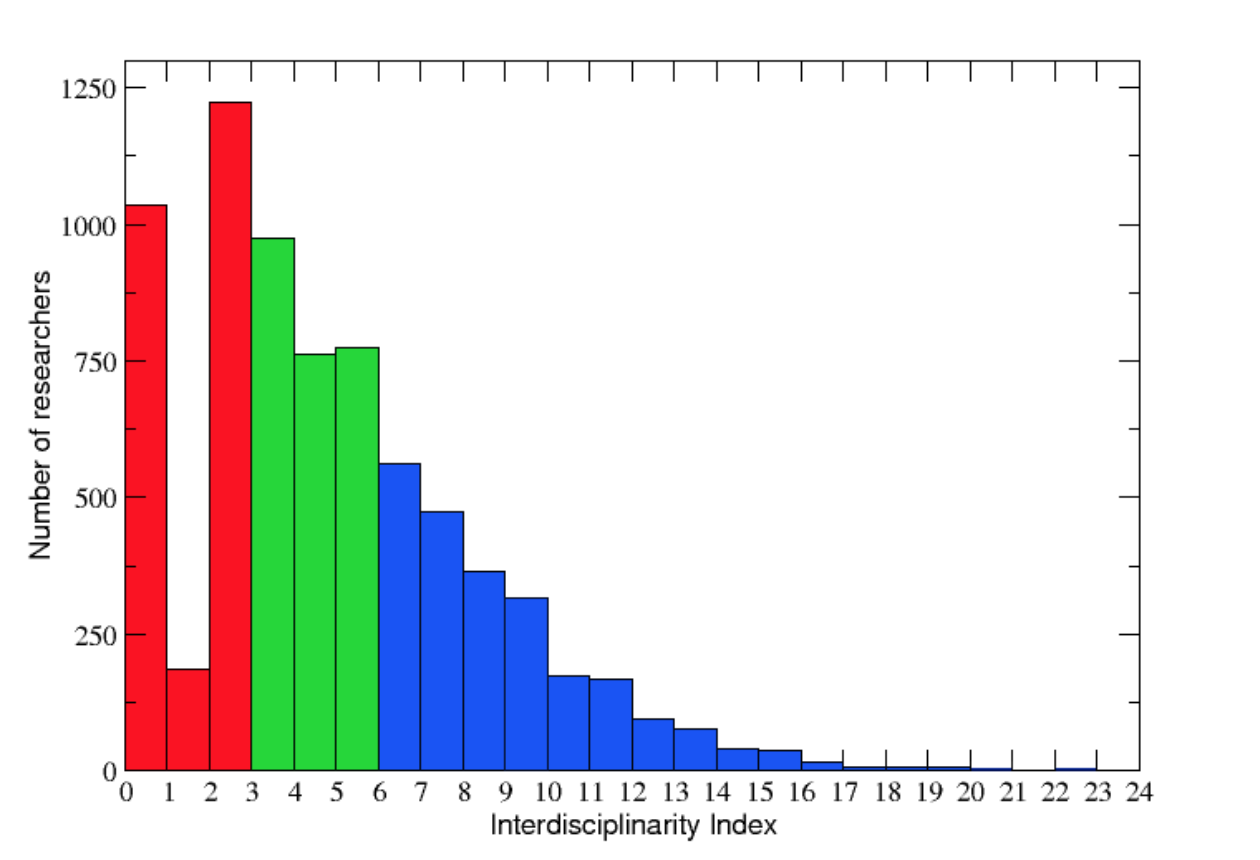}
			\caption[]{
				\label{generalQ}
				(Left Panel) An example of calculation of the interdisciplinarity index for an imaginary author $A_i$ who published 6 papers. (Right Panel) Histogram of the interdisciplinarity index $I^{APS}$ for the $N = 7303$ researchers interested by our study. The three different interdisciplinarity levels are represented with different colors: red (level 1), green (level 2) and blue (level 3). The bar for $I^{APS}$ between $j$ and $j+1$ represents the number of researchers with $I^{APS}\in]j,j+1]$. In particular, the first two bars contains only researchers with $I^{APS}=1$ and $I^{APS}=2$, respectively.
			}
		\end{center}
	\end{figure}
	
	\subsubsection{S1.1 Researchers Classification}
	
	Having at our disposal the PACS coded areas of all the papers, we may use them to define an index that helps us to quantify the variety of disciplines (areas) interested by the scientific production of any researcher. This variety is two-fold: a researcher may explore many different areas one by one, i.e. producing on many different PACS codes through papers with assigned only one code at a time; or she may explore few different areas but jointly, i.e. producing papers having more codes assigned together. In other words, a researcher's production can be interdisciplinary either because of the total number of areas that it interested, or because of the average number of areas jointly interested in one of its typical paper. As it is going to be evident, apart from an obvious constraint, these two degrees of interdisciplinarity are independent of each other. This observation led us to define an interdisciplinary index $I^{APS}_{k}$ for the researcher $A_k$ as
	\begin{equation*}
	I^{APS}_{k} = D^{APS}_{k} \times d^{APS}_{k} 
	\end{equation*}	
where $d^{APS}_{k}$ is the average number of different PACS codes jointly present in each paper of the considered author and $D^{APS}_{k}\in[1,10]$ is the total number of different PACS codes present in all the papers of the same author. One can also imagine to assign to $A_k$ an array $\vec{D}^{APS}_{k}$ containing all the $D^{APS}_{k}$ PACS numbers present in her papers. The constraint mentioned above is the mere condition $d^{APS}_{k}$ $\leq$ $D^{APS}_{k}$ for any $k$. In fact, the maximum number of PACS codes assignable to a paper is five, so, at least in principle, the maximum value of $I^{APS}$ is 50, with $d^{APS}=5$ and $D^{APS}=10$. In practice, for our data set, the maximum value found for $I$ is 23, with $d^{APS}=3.286$ and $D^{APS}=7$.
In the left panel of Fig.10 an example of calculation of the interdisciplinarity index for a hypothetic author $A_i$ is presented. This author has published $P_i=6$ papers, each one with different PACS numbers (1-6, 4, 1, 4-8, 6-8-1, 1, respectively). The corresponding PACS array is thus $\vec{D}^{APS}_{i}=[1 4 6 8]$,  $D^{APS}_i=4$ and $d^{APS}_{i}=1.66$. Therefore, her interdisciplinarity index will be $I^{APS}_{i}=6.64$.      

	Once the interdisciplinarity index has been calculated for each researcher, we have distributed all the 7303 authors - resulted from the filtering procedure explained below - into three groups of different interdisciplinarity level (see right panel of Fig.10):	
	\begin{itemize}
\item Level 1 ($L^{APS}_1$):\, $1$ $\leq$ $I^{APS}_{k}$ $\leq$ $3$ \,\,\,($N_1=2445$ researchers of low interdisciplinarity level)
\item Level 2 ($L^{APS}_2$):\, $3$ $<$ $I^{APS}_{k}$ $\leq$ $6$ \,\,\,($N_2=2511$ researchers of medium interdisciplinarity level)
\item Level 3 ($L^{APS}_3$):\, $I^{APS}_{k}$ $>$ $6$ \,\,\,($N_3=2347$ researchers of high interdisciplinarity level)
	\end{itemize}	
The separation values between the levels have been chosen to have the three groups with comparable sizes and, for the set of researchers used here, the best values came out to be 3 and 6, if we want them as easy-to-remind integer numbers. To note that for the level 1, because of the condition $d^{APS}_{k}$ $\leq$ $D^{APS}_{k}$, the index $I^{APS}_{k}$ cannot take value in the open interval (1,2).
		
				\begin{figure}[t]
		\begin{center}
			\includegraphics[width=.5\linewidth]
			{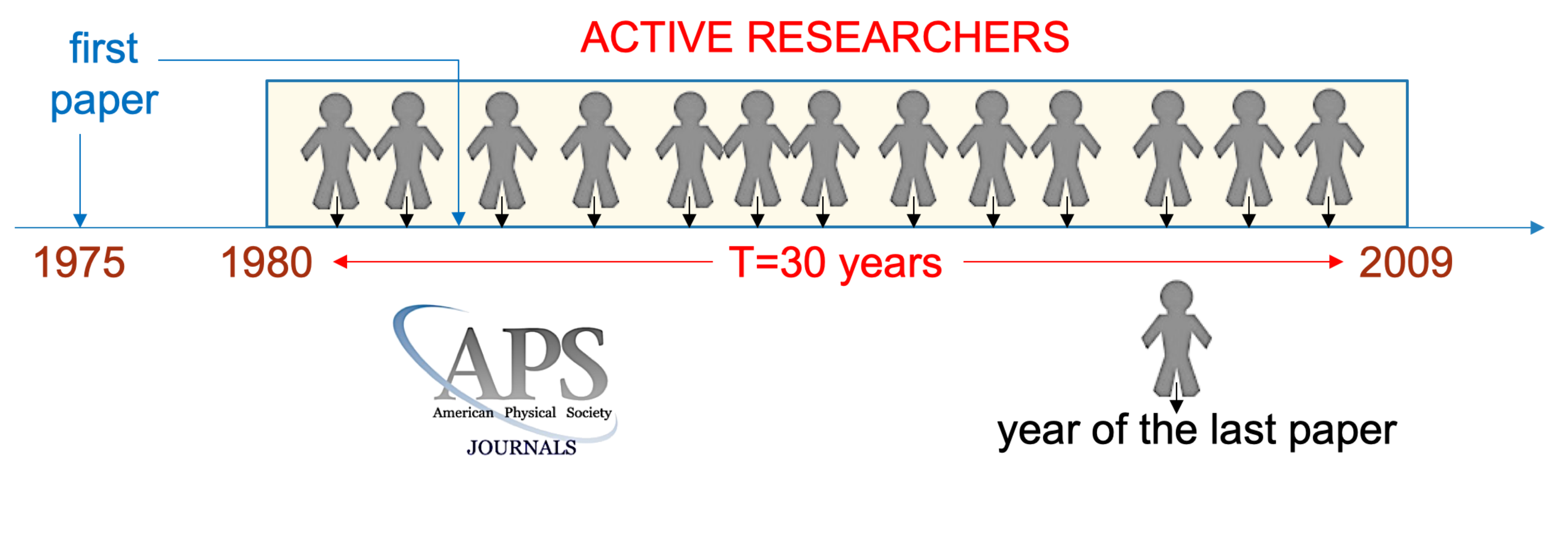}
			\includegraphics[width=.4\linewidth]
			{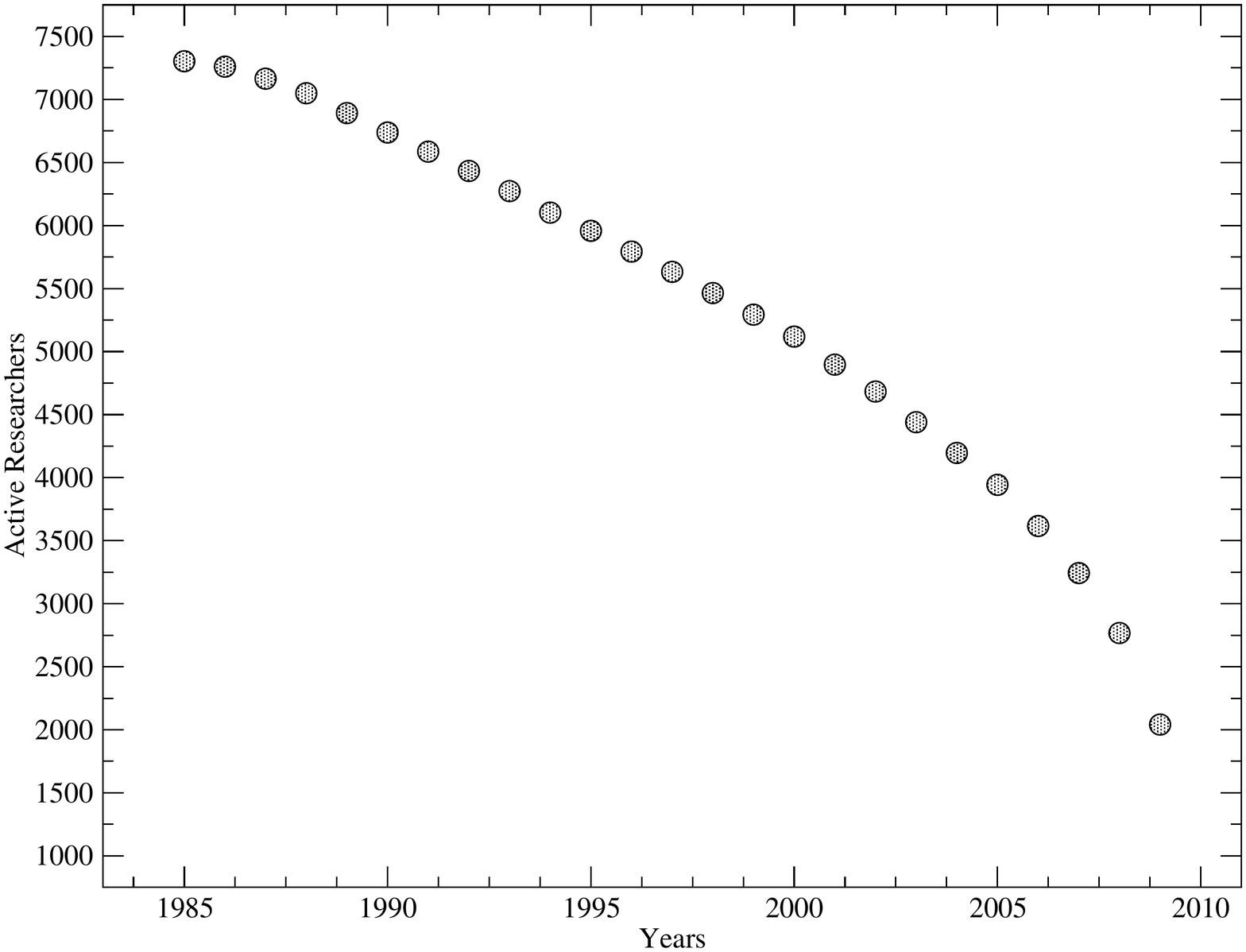}
			\caption[]{
				\label{generalQ}
				(Left Panel) The active researchers considered in the APS data set analysis, see text. (Right Panel) Time evolution, year by year, of the number of still active researchers. A linear decrease is found from 1987 to 2002, with 165 leaving researchers a year, on average. After 2002 a kind of cut off acts, maybe due to the their ages. The 28\% of them is still active at the end of the thirty years. 
			}
		\end{center}
	\end{figure}	
		
	The 7303 researchers on which we have conducted our analysis are the remaining ones of a filtering procedure conceived to study appropriately the researchers' careers over a period of thirty years, from 01/01/1980 to 31/12/2009. The first requirement of the filtering is that a researcher must have produced her first paper in the period ranging from 01/01/1975 to 31/12/1985 (see the left panel of Fig.11). This ensures that all the researchers in the set started their careers in a quite short period, so avoiding that the possible premature end of the production activity of a researcher is due to her age. In this way, unless one started to produce in old age, that is a pretty remote possibility, all the researchers in the set have comparable ages. Moreover, the PACS classification was implemented from 1975 onwards, enabling us to refer only to papers published starting from that year. The second requirement is that a researcher must have produced a minimum number of (PACS classified) papers, that we chose to be 3. The third, last, requirement is related to the way in which the raw APS database at our disposal has been cleaned (extensively explained in the specific section). 	
	
	Briefly, at each author's name has been given an author identification code and the same code has been assigned to different names if they were similar enough. We refer to the authors' name associated with the same author code as aliases of that author. We ruled out those author codes with more than one alias associated to it. We realized, indeed, that not enough rarely happened that two aliases referred to two actually different authors (with similar names, unfortunately), leading us to overestimate the productivity and the impact of the unique author code which they were assigned to. These three requirements filtered the database leaving us with 7303 initial author codes, corresponding to the 7303 actually different researchers on which we have performed our analysis.
	
	Looking at the last published paper by each researcher, apart of a late cut off, an approximately linear decrease in time of the number of active researchers came out. Starting with all the 7303 researchers active in 1985, we end up with 2041 of them still active in 2009 (Fig.11, right panel).

	\subsubsection{S1.2 Scientific Impact Analysis}
	
	The scientific production in the period 1980-2009 of the 7303 selected researchers consists of 89949 (PACS classified) papers. These are distributed in a slightly different way over the three defined classes of interdisciplinarity, see the left panel of Fig.12. In all of them one can note long tails of a few dozen of researchers with an exceptional productivity, but in general interdisciplinarity seem to have a positive influence on the average productivity of a scientist. Some examples of the increase in the scientific production during single excellent careers for the three classes is shown in the right panel of Fig.12, where the cumulated number of papers is reported as function of time. 
	
	\begin{figure}[t]
	\begin{center}
			\includegraphics[width=.44\linewidth]
			{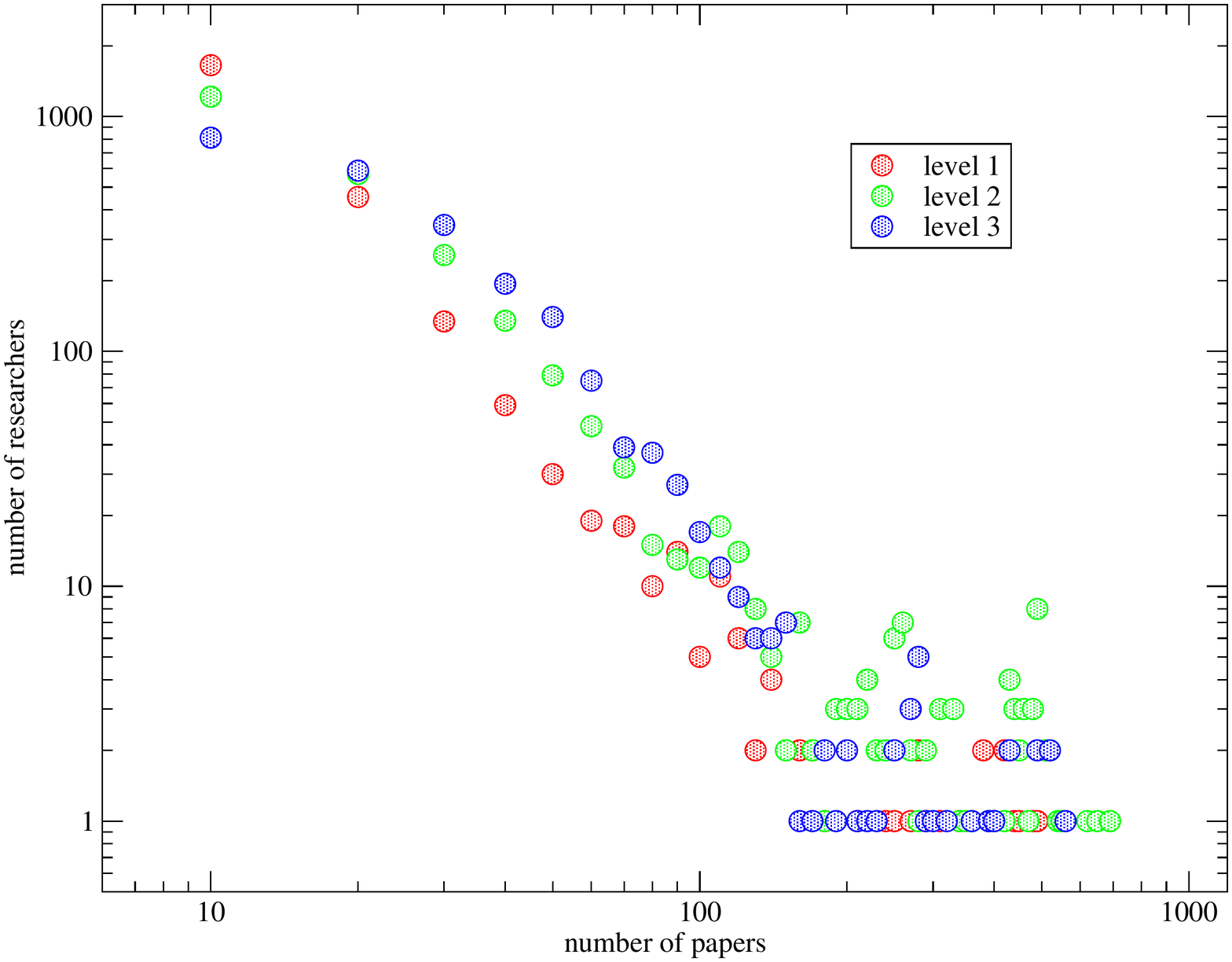}
			\includegraphics[width=.43\linewidth]
			{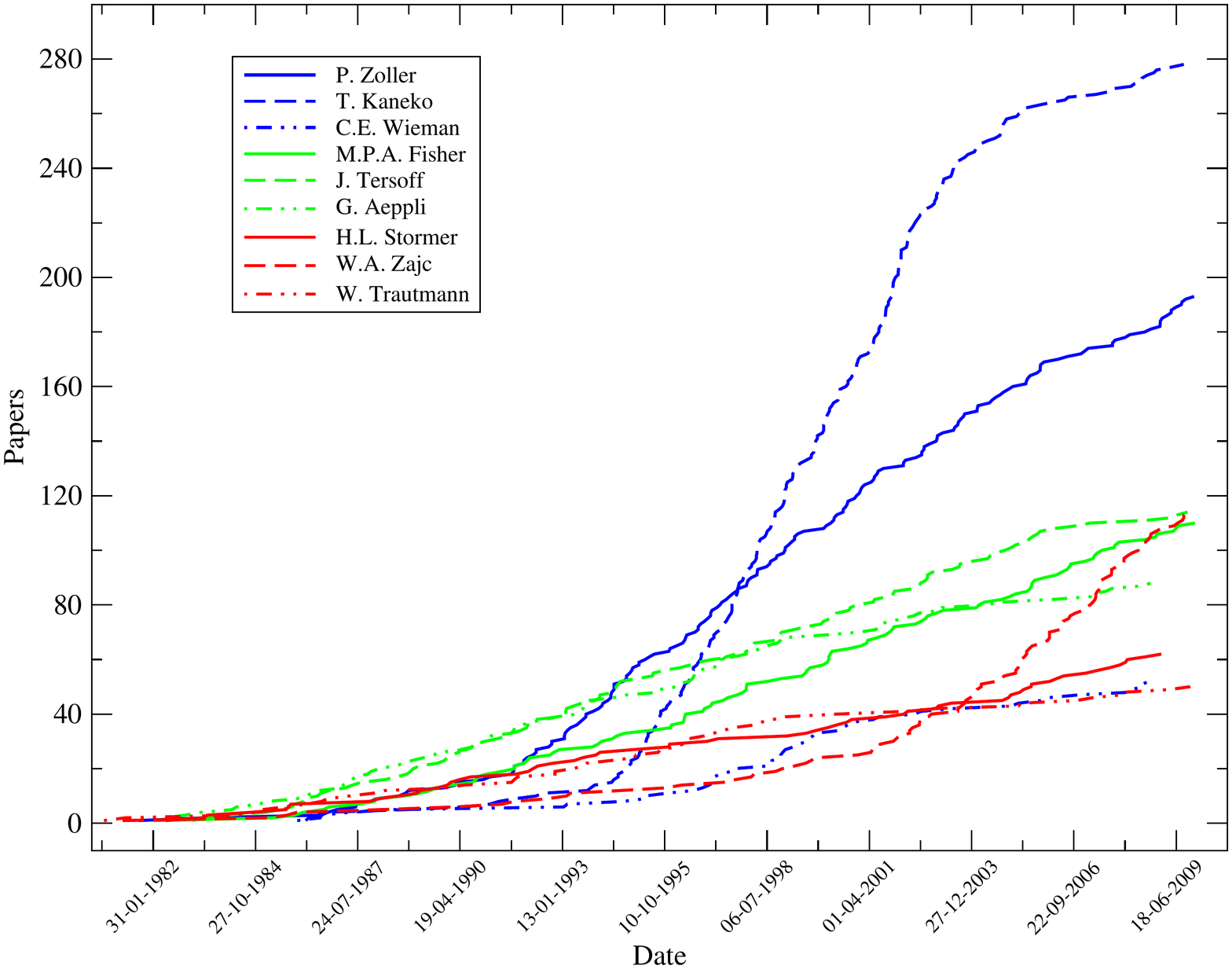}
	\caption[]{
				\label{generalQ}
				(Left Panel) Papers distribution for the three defined classes of interdisciplinarity, each represented with a different color: red (level 1), green (level 2) and blue (level 3). A tail of scarse statistics starts for numbers of researchers with more than about 150 published papers. (Right Panel) Examples of scientific production in some excellent careers for the three interdisciplinarity classes. 
			}
	\end{center}
	\end{figure}

	\begin{table}[]
	\begin{tabular}{c|c|c|c|c|}
	\cline{2-5}
	                              & \,authors\, & \,papers\, & \,PpA\, & \,avg.\,PpA\,(st.\,dev.)\, \\ 	\hline
	\multicolumn{1}{|c|}{level 1} & 2445        & 18832  & 7.70                                       & 15.38 (37.22)                                             \\ \hline
	\multicolumn{1}{|c|}{level 2} & 2511        & 35892  & 14.29                                      & 29.35 (67.18)                                             \\ \hline
	\multicolumn{1}{|c|}{level 3} & 2347        & 50947  & 21.71                                      & 27.30 (42.26)                                             \\ \hline
	\end{tabular}	\caption[]{
				\label{generalQ}
				Statistical indicators of the 89949 published papers over the three defined classes of interdisciplinarity. A paper is counted in more than one class if it is coauthored by researchers belonging to different classes, so the sum of the reported numbers of papers exceeds 89949. A positive correlation between scientific production and interdisciplinarity level is found: the number of papers per researcher (PpA = papers/authors) increases quite strongly as the interdisciplinarity level grows.}
	\end{table}
				\
	
\begin{table}[H]
\begin{tabular}{cc|c|c|c|c|c|c|c|c|c|c|}
\cline{3-12}    &             & \multicolumn{10}{c|}{PACS Area}                                                                                                                                                                                                                                                                                                                                                                                                                                                                                                                                                                          \\ \cline{3-12} &             & 00                                                        & 10                                                       & 20                                                       & 30                                                       & 40                                                       & 50                                                      & 60                                                        & 70                                                        & 80                                                       & 90                                                      \\ \hline
\multicolumn{1}{|c|}{\multirow{2}{*}{\begin{tabular}[c]{@{}c@{}}Level\\ 1\end{tabular}}} & papers      & \begin{tabular}[c]{@{}c@{}}609\\ (3.23\%)\end{tabular}    & \begin{tabular}[c]{@{}c@{}}4892\\ (25.98\%)\end{tabular} & \begin{tabular}[c]{@{}c@{}}5989\\ (31.80\%)\end{tabular} & \begin{tabular}[c]{@{}c@{}}1488\\ (7.90\%)\end{tabular}  & \begin{tabular}[c]{@{}c@{}}305\\ (1.62\%)\end{tabular}   & \begin{tabular}[c]{@{}c@{}}720\\ (3.82\%)\end{tabular}  & \begin{tabular}[c]{@{}c@{}}1518\\ (8.06\%)\end{tabular}   & \begin{tabular}[c]{@{}c@{}}5232\\ (27.78\%)\end{tabular}  & \begin{tabular}[c]{@{}c@{}}93\\ (0.49\%)\end{tabular}    & \begin{tabular}[c]{@{}c@{}}236\\ (1.25\%)\end{tabular}  \\ \cline{2-12} 
\multicolumn{1}{|c|}{}                                                                   & researchers & \begin{tabular}[c]{@{}c@{}}231\\ (9.45\%)\end{tabular}    & \begin{tabular}[c]{@{}c@{}}782 \\ (31.98\%)\end{tabular} & \begin{tabular}[c]{@{}c@{}}811 \\ (33.17\%)\end{tabular} & \begin{tabular}[c]{@{}c@{}}277\\ (11.33\%)\end{tabular}  & \begin{tabular}[c]{@{}c@{}}128\\ (5.24\%)\end{tabular}   & \begin{tabular}[c]{@{}c@{}}197\\ (8.06\%)\end{tabular}  & \begin{tabular}[c]{@{}c@{}}475\\ (19.43\%)\end{tabular}   & \begin{tabular}[c]{@{}c@{}}744 \\ (30.43\%)\end{tabular}  & \begin{tabular}[c]{@{}c@{}}68\\ (2.78\%)\end{tabular}    & \begin{tabular}[c]{@{}c@{}}98 \\ (4.01\%)\end{tabular}  \\ \hline
\multicolumn{1}{|c|}{\multirow{2}{*}{\begin{tabular}[c]{@{}c@{}}Level\\ 2\end{tabular}}} & papers      & \begin{tabular}[c]{@{}c@{}}3244\\ (9.04\%)\end{tabular}   & \begin{tabular}[c]{@{}c@{}}7466\\ (20.80\%)\end{tabular} & \begin{tabular}[c]{@{}c@{}}6703\\ (18.68\%)\end{tabular} & \begin{tabular}[c]{@{}c@{}}3006\\ (8.38\%)\end{tabular}  & \begin{tabular}[c]{@{}c@{}}1715\\ (4.78\%)\end{tabular}  & \begin{tabular}[c]{@{}c@{}}1064\\ (2.96\%)\end{tabular} & \begin{tabular}[c]{@{}c@{}}6101\\ (17.00\%)\end{tabular}  & \begin{tabular}[c]{@{}c@{}}15013\\ (41.83\%)\end{tabular} & \begin{tabular}[c]{@{}c@{}}1361\\ (3.79\%)\end{tabular}  & \begin{tabular}[c]{@{}c@{}}1121\\ (3.12\%)\end{tabular} \\ \cline{2-12} 
\multicolumn{1}{|c|}{}                                                                   & researchers & \begin{tabular}[c]{@{}c@{}}1032\\ (41.10\%)\end{tabular}  & \begin{tabular}[c]{@{}c@{}}849\\ (33.81\%)\end{tabular}  & \begin{tabular}[c]{@{}c@{}}794\\ (31.62\%)\end{tabular}  & \begin{tabular}[c]{@{}c@{}}685\\ (27.28\%)\end{tabular}  & \begin{tabular}[c]{@{}c@{}}528\\ (21.03\%)\end{tabular}  & \begin{tabular}[c]{@{}c@{}}220\\ (8.76\%)\end{tabular}  & \begin{tabular}[c]{@{}c@{}}1213\\ (48.31\%)\end{tabular}  & \begin{tabular}[c]{@{}c@{}}1317\\ (52.45\%)\end{tabular}  & \begin{tabular}[c]{@{}c@{}}696\\ (27.72\%)\end{tabular}  & \begin{tabular}[c]{@{}c@{}}367\\ (14.62\%)\end{tabular} \\ \hline
\multicolumn{1}{|c|}{\multirow{2}{*}{\begin{tabular}[c]{@{}c@{}}Level\\ 3\end{tabular}}} & papers      & \begin{tabular}[c]{@{}c@{}}12397\\ (24.33\%)\end{tabular} & \begin{tabular}[c]{@{}c@{}}6056\\ (11.89\%)\end{tabular} & \begin{tabular}[c]{@{}c@{}}4700\\ (9.23\%)\end{tabular}  & \begin{tabular}[c]{@{}c@{}}6430\\ (12.62\%)\end{tabular} & \begin{tabular}[c]{@{}c@{}}7265\\ (14.26\%)\end{tabular} & \begin{tabular}[c]{@{}c@{}}1813\\ (3.56\%)\end{tabular} & \begin{tabular}[c]{@{}c@{}}14159\\ (27.79\%)\end{tabular} & \begin{tabular}[c]{@{}c@{}}21461\\ (42.12\%)\end{tabular} & \begin{tabular}[c]{@{}c@{}}5612\\ (11.02\%)\end{tabular} & \begin{tabular}[c]{@{}c@{}}1867\\ (3.66\%)\end{tabular} \\ \cline{2-12} 
\multicolumn{1}{|c|}{}                                                                   & researchers & \begin{tabular}[c]{@{}c@{}}1705\\ (72.65\%)\end{tabular}  & \begin{tabular}[c]{@{}c@{}}743\\ (31.66\%)\end{tabular}  & \begin{tabular}[c]{@{}c@{}}699\\ (29.78\%)\end{tabular}  & \begin{tabular}[c]{@{}c@{}}1216\\ (51.81\%)\end{tabular} & \begin{tabular}[c]{@{}c@{}}1348\\ (57.44\%)\end{tabular} & \begin{tabular}[c]{@{}c@{}}503\\ (21.43\%)\end{tabular} & \begin{tabular}[c]{@{}c@{}}1790\\ (76.27\%)\end{tabular}  & \begin{tabular}[c]{@{}c@{}}1790\\ (76.27\%)\end{tabular}  & \begin{tabular}[c]{@{}c@{}}1447\\ (61.65\%)\end{tabular} & \begin{tabular}[c]{@{}c@{}}460\\ (19.60\%)\end{tabular} \\ \hline
\end{tabular}
	\caption[]{
				\label{generalQ}
				Distribution of the researchers of each interdisciplinarity level and their papers through the ten PACS coded areas. 
			}
	\end{table}
	
A confirm of the positive correlation between scientific production and interdisciplinarity level is shown in Table III. Comparing the number of papers per author (PpA) and the (real) average number of papers per author (avg. PpA), we also find a stronger presence of coauthoring in the level 1 and level 2 classes than in the level 3 class. This is due mainly to the fact that a lower percentage of researchers of the level 3 class participated to large scientific collaboration, respect to the other two classes. 
	
	\begin{figure}
	\begin{center}
			\includegraphics[width=.44\linewidth]
			{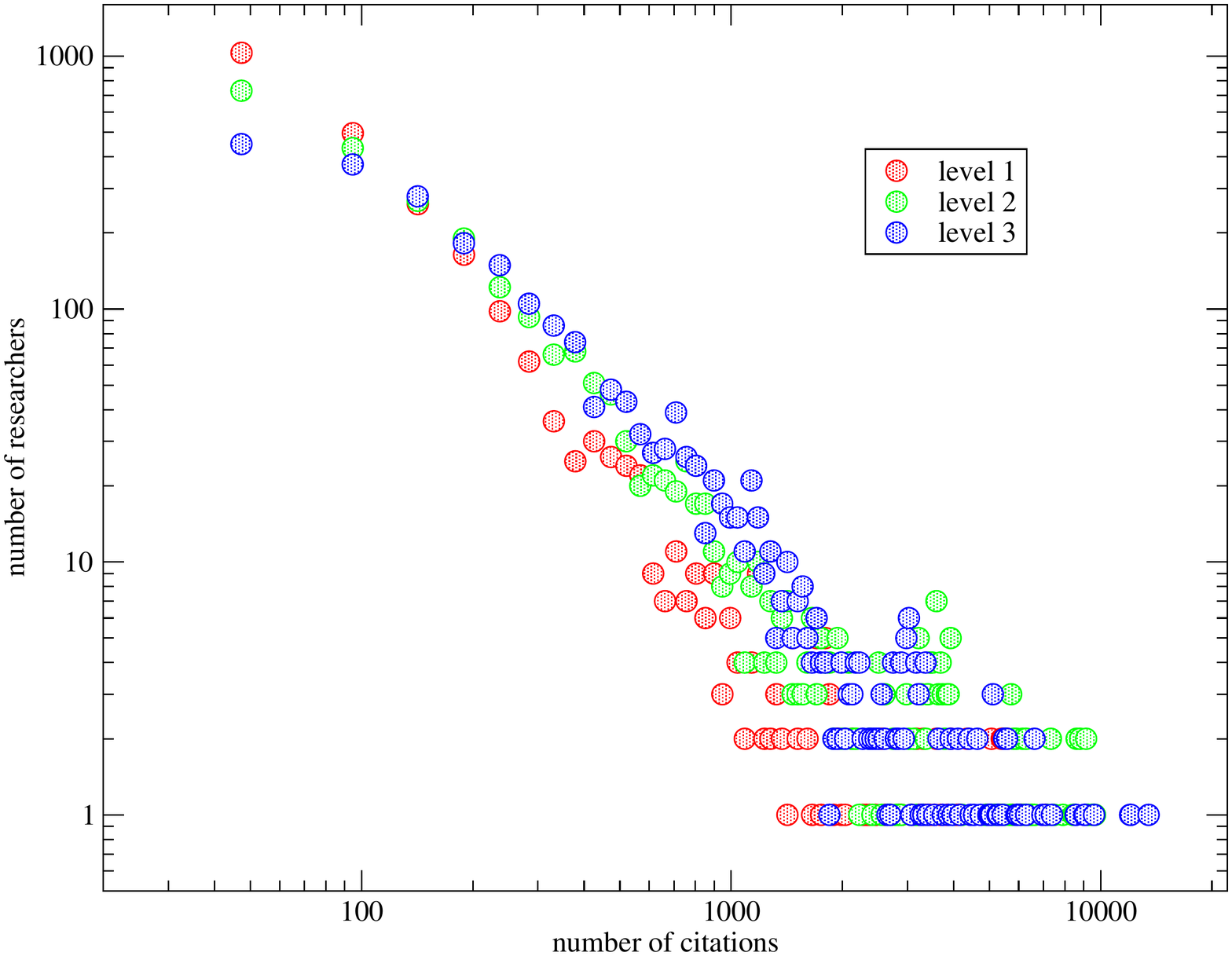}
			\includegraphics[width=.43\linewidth]
			{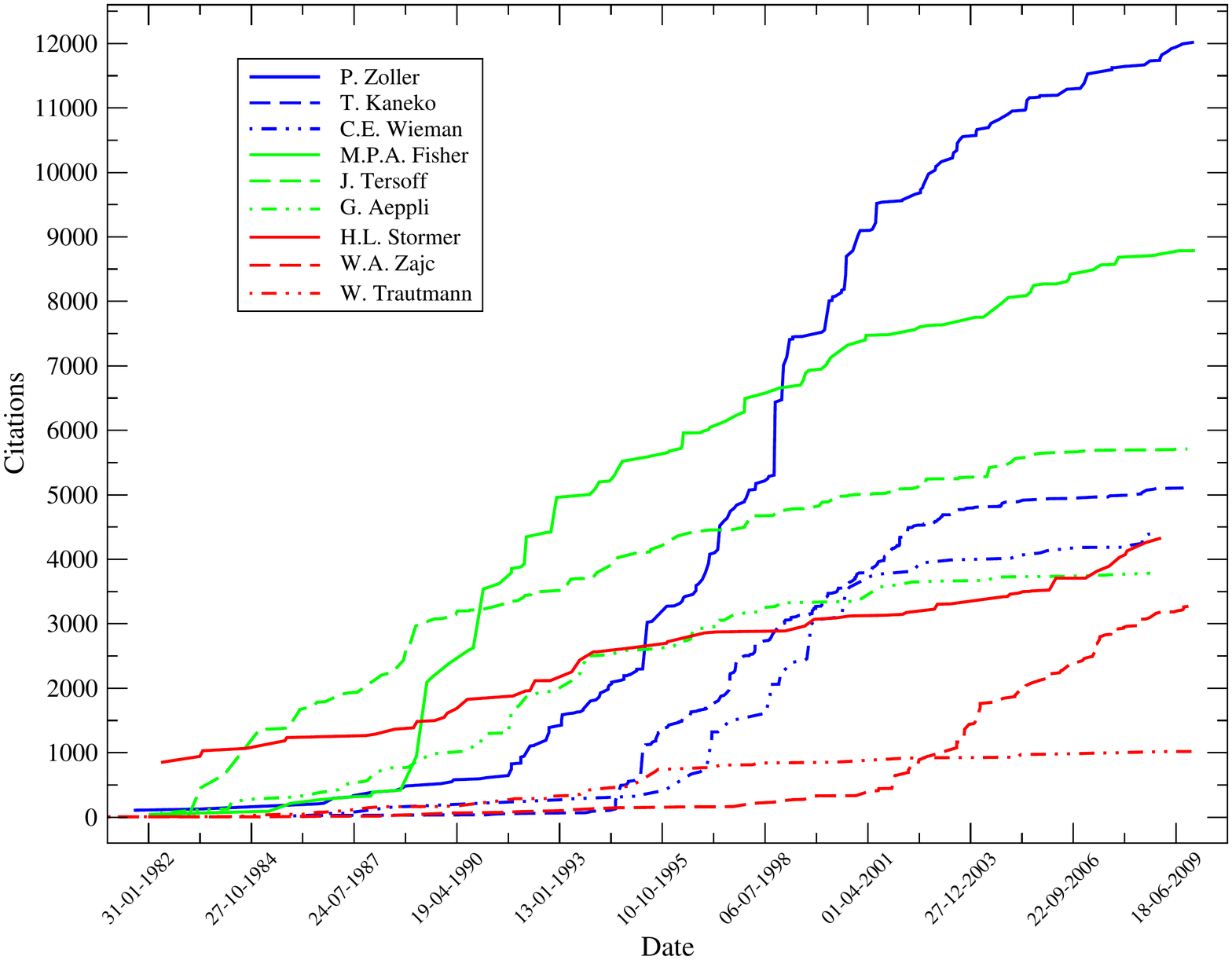}
	\caption[]{
				\label{generalQ}
				(Left Panel) Citations distribution for the three defined classes of interdisciplinarity, each represented with a different color: red (level 1), green (level 2) and blue (level 3). (Right Panel) The same careers shown in Fig.12 are here addressed in terms of time evolution of scientific impact. 
			}
	\end{center}
	\end{figure}
	
\begin{table}[]
	\begin{tabular}{c|c|c|c|c|c|}
	\cline{2-6}
	                              & \,authors\, & \,papers\, & \,citations\, & \,CpA\, & \,avg.\,CpA\,(st.\,dev.)\, \\ \hline
	\multicolumn{1}{|c|}{level 1} & 2445        & 18832  & 230448   & 94.25                        & 217.52  (598.48)                      \\ \hline
	\multicolumn{1}{|c|}{level 2} & 2511        & 35892  & 515635  & 205.35                       & 458.44  (1121.24)                     \\ \hline
	\multicolumn{1}{|c|}{level 3} & 2347        & 50947  & 843292  & 359.31                        & 479.07  (997.75)                     \\ \hline
	\end{tabular}
	\caption[]{
				\label{generalQ}
				Statistical indicators of the citations received by the authors and their papers for each of the three defined classes of interdisciplinarity. All these citations divide slightly differently for each class (Fig.13, left panel). A positive correlation between scientific impact, in terms of citations received, and interdisciplinarity level is found: the number of citations per author (CpA = citations/authors) raises as the interdisciplinarity level increases.
			}
	\end{table} 
	

By looking minutely at their production one finds out that all of them did research in the areas of particle and nuclear physics. More precisely, these researchers took part in large scientific experiments (e.g. BABAR, CLEO, CDF collaborations) during the 2000s. These large collaborations of hundreds of scientists ensure to the participants high rates of scientific productivity of even 60/70 published papers a year, an unachievable goal for the small research groups working in other areas. As proved by the composition of the three interdisciplinarity classes in terms of the ten PACS coded areas - see Table IV - most of the researchers in our set who are involved in these large collaborations belong to the level 2 class, justifying the heavier tail found for this class compared to those found for the other two classes (Fig.12).

One easily notes that these indicators clearly underestimate the real productivity of the researchers, but it must be kept in mind that they refer only to (PACS coded) publications on APS and that the actual number of researchers decreased over the thirty years, as shown in Fig.11.

The 89949 (PACS classified) published papers of the set received a total of 1329374 citations within the APS system in the period 1980-2009. From the point of view of the 7303 researchers, considered as independent, they received a total of 2807368 citations in the same period. All these citations divide similarly among the researchers of each of the three interdisciplinarity classes, as shown in the left panel of Fig.13. Also in this case, as previously shown for the papers production, we found a positive correlation between scientific impact, in terms of citations received, and interdisciplinarity level (Table V). Finally, in the right panel of Fig.13, the increase in the number of citations cumulated by the same excellent careers considered in Fig.12 is reported as function of time. Notice that not necessarily the best score in terms of published papers does imply the best score in terms of scientific impact and vice-versa.

\begin{figure}
\begin{center}
\includegraphics[width=5.0 in,angle=0]{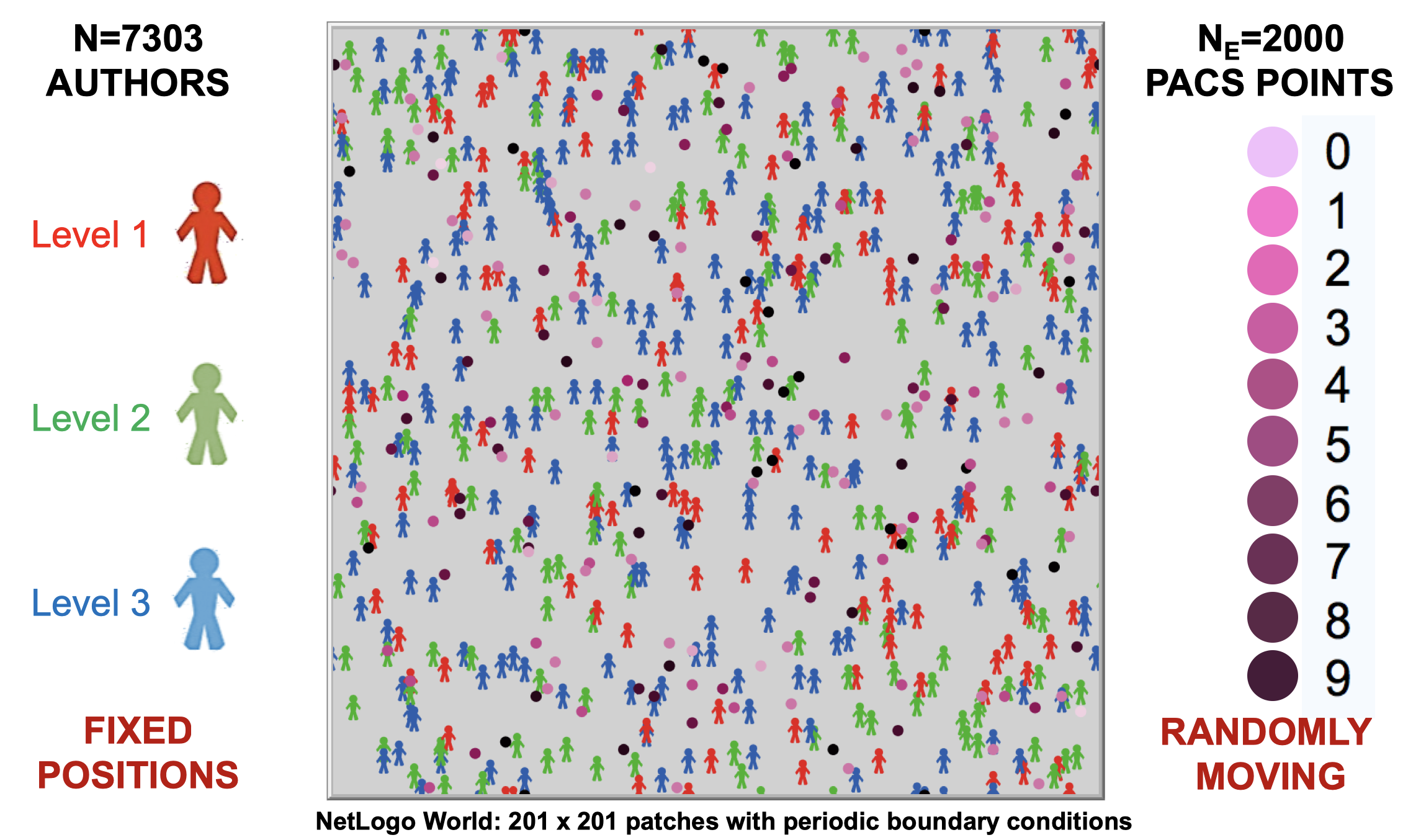}
\caption{\small 
An example of initial setup for our simulations.        
}
\label{world} 
\end{center}
\end{figure}
	
As a final curiosity, apart from these excellences, let us see some other authors names belonging to the three interdisciplinarity groups of our data set. 
In particular, in the $L^{APS}_1$ group one find mainly scientist who have been working in nuclear physics, like W. Alberico, U. Lynen, Y.T. Oganessian, W. Trautmann. 
On the other hand, in the $L^{APS}_2$ group one can find scientists who worked in various fields, from chaos theory to gravitational waves, or from quantum information to cosmology, as for example C. Grebogi, D. Deutsch, K. Wilson, J.E. Jaffe, L. Smolin, P.C.W. Davies, G. Pizzella. 
Finally, in the most interdisciplinar $L^{APS}_3$ group, one finds mainly statistical or condensed matter physicists, scientists involved in complex networks and dynamical systems, and also cosmologists or experts of string theory with broad views (P. Bak, A. Coniglio, K. Kaneko, M. Mezard, S. Havlin, D. Sornette, G. Parisi, J. Barrow and B. Greene).


\subsection{S2. The agent-based model}

Let us address, now, some details about the agent-based model with which we were able to successful replicate the stylized facts of the APS data set. The model was realized within NetLogo, a very powerful multi-agent programmable environment particularly suitable for the the simulation of the dynamical behavior of complex systems \cite{netlogo}.

\subsubsection{S2.1 Initial setup of the model}

In Fig.\ref{world} we show the NetLogo "world" as it appears at the beginning of a generic simulation. It is a squared metric space, with a size of $201 \times 201$ patches, where the various agents live and move. Randomly distributed around the world are visible the two main categories of agents of our model: $N$ researchers, with a person-like shape, and $N_E$ PACS event-points, with a point-like shape. Both these agent's types are active elements of the environment, able of interact one among each other.   

\begin{figure}[t]
\begin{center}
\includegraphics[width=3.4in,angle=0]{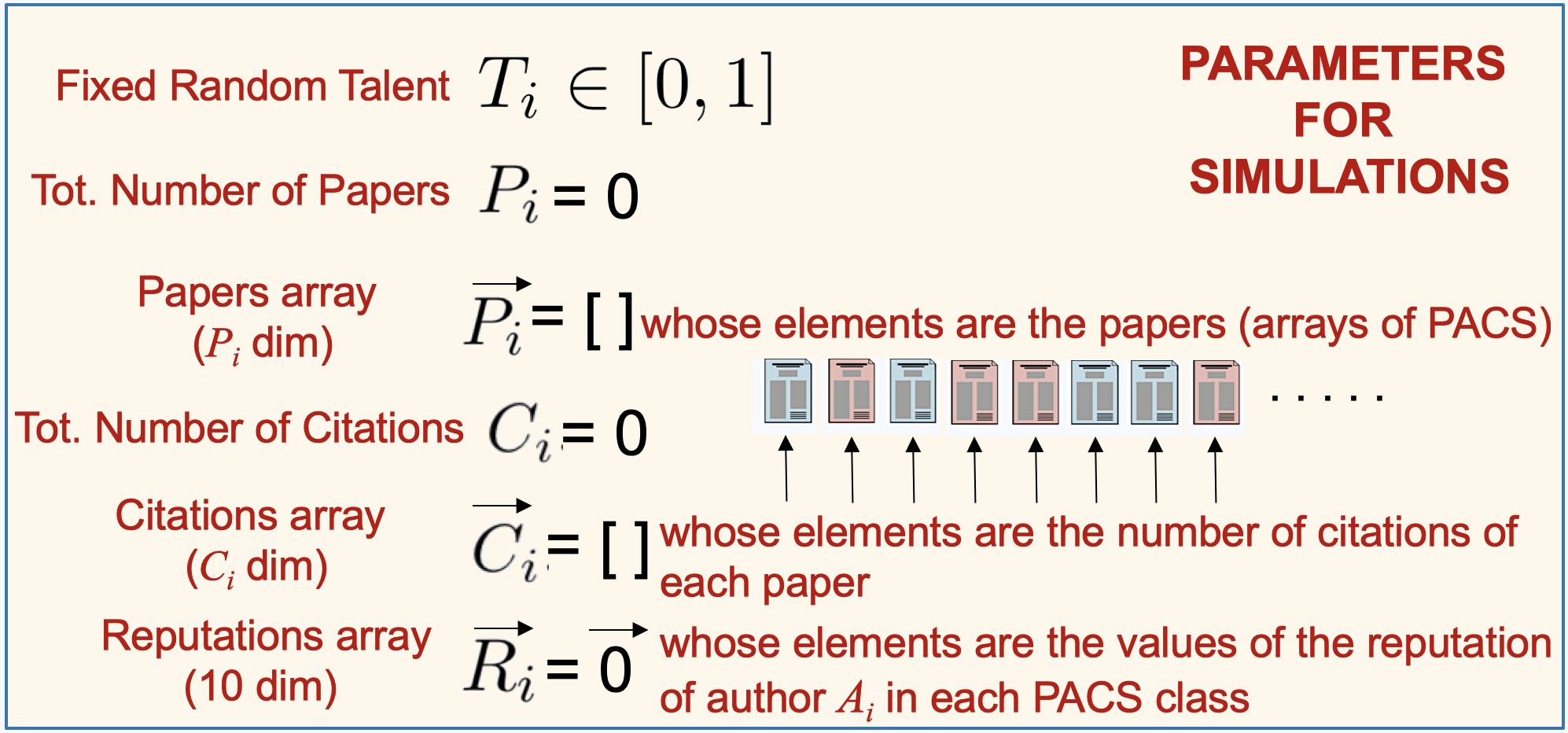} 
\includegraphics[width=3in,angle=0]{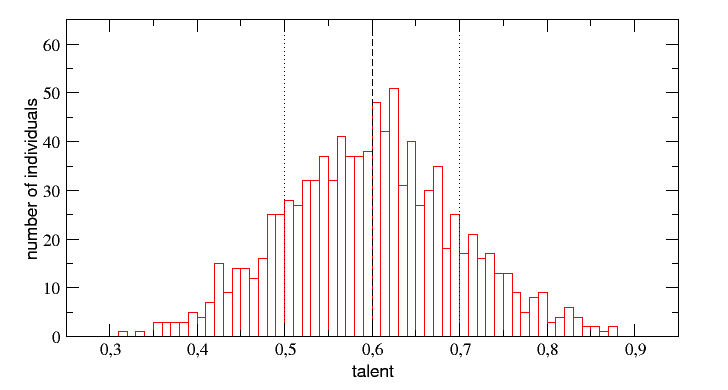} 
\caption{\small 
(Left Panel) Individual parameters which characterize each single simulated author $A_i$. (Right Panel) Normal distribution of talent among the agents, with mean $m_T=0.6$ (indicated by a dashed vertical line) and standard deviation $\sigma_T=0.1$ (the values $m_T \pm \sigma_T$ are indicated by two dotted vertical lines). This distribution does not change during a single simulation run.        
}
\label{talent} 
\end{center}
\end{figure}

In the figure we represent only $N=500$ individuals for a better visualization, but in all the simulations we consider all the $N=7303$ active researchers, as in the APS data set. These researchers do no move during a simulation and are divided into the three groups $L^{APS} _1$, $L^{APS}_2$ and $L^{APS}_3$ according with the real values of their interdisciplinary index $I^{APS}_i = D^{APS}_i \times d^{APS}_i$. Therefore, we will find $N_1$ individuals in the group $L^{APS}_1$ (in red), $N_2$ in the group $L^{APS}_2$ (in green) and $N_3$ in the group $L^{APS}_3$ (in blue). During a single simulation run, we will let these researchers to publish papers and collect citations with a periodicity of $t=1$ year and for a total time interval of $t_{max}=30$ years, in analogy with the real time period addressed in the APS data set. A first evident approximation of the model is the fact that we will keep the total number of active authors constant during the $30$ years, while we know that their number do decrease, as shown in Fig.11. This will imply an overestimation of the total number of published papers of several authors, but - as we have already stated - we are interested to capture the main stylized facts of the APS data set not the single details (which, of course, would be impossible to reproduce).     

Each simulated author $A_i$ is characterized not only by the variables $I^{APS}_i$, $D^{APS}_i$, $\vec{D}^{APS}_i$ and $d^{APS}_i$ ($i=1,...,N$), which are read from the APS data set, but also by other individual parameters shown in the left panel of Fig.S6 and described in the MP. In particular, to each researcher is assigned a fixed talent $T_i \in [0,1]$ (intelligence, skill, ...) randomly extracted at the beginning of each simulation run from a truncated Normal distribution with a mean $m_T=0.6$ and a standard deviation $\sigma_T=0.1$ (see the right panel of Fig.\ref{talent}). All the other individual parameters start from a null value at $t=0$ and increase in time during the simulation following opportune dynamical rules.  

As we will show in the next subsection, other global parameters need to be introduced in the model and calibrated through the comparison with the real APS data.     

\begin{figure}[b]
\begin{center}
\includegraphics[width=3.2 in,angle=0]{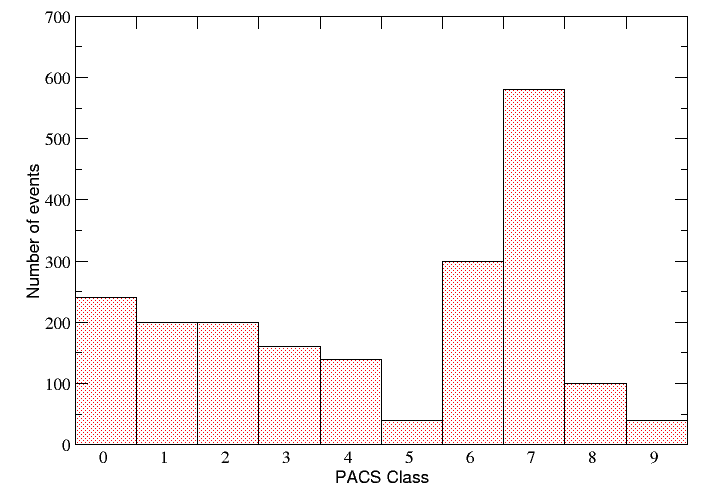}
\includegraphics[width=2.54 in,angle=0]{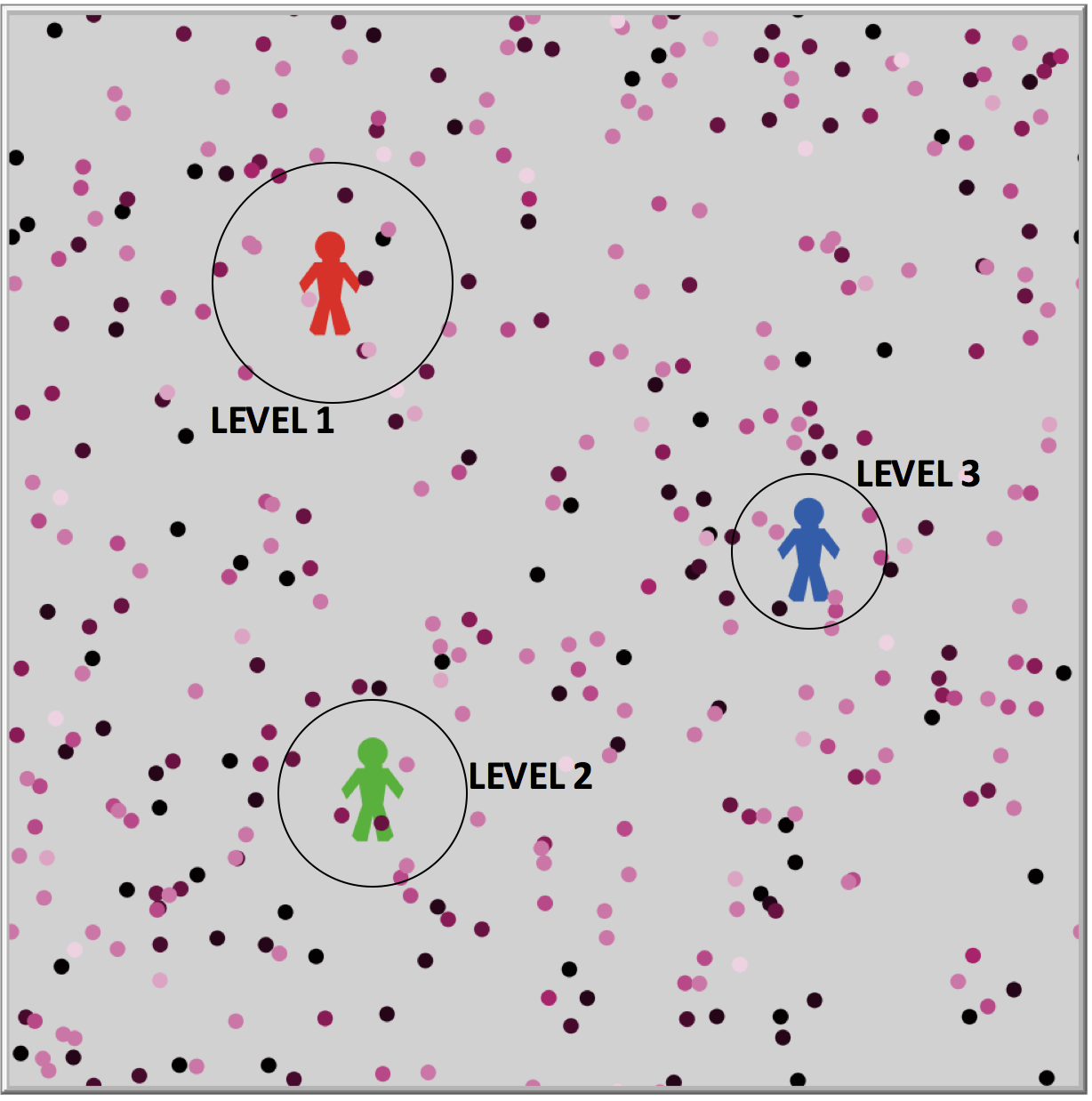}
\caption{\small 
Left panel: A histogram showing the number of event points for each PACS class, over a total of $N_E=2000$, according to its relative percentage abundance in the APS dataset.      
Right panel: A zoom from Fig.\ref{world}, where only three researchers, each belonging to one of the three interdisciplinarity levels, are reported with their colors: red (level 1), green (level 2) and blue (level 3). Around them, some moving events are visible, represented as points of different colors selected from a magenta scale. Each color corresponds to a given PACS class of the APS data set, numbered from 0 (darkest) to 9 (brightest), as also shown in Fig.S5. The relative percentage of event points of each class is different and corresponds to the real one. Around each of the three researchers, the corresponding sensitivity circle is also visible, whose radius decreases by increasing the interdisciplinarity level (see text).                
}
\label{world-zoom} 
\end{center}
\end{figure}

\subsubsection{S2.2 Calibration of the model}

The first global parameters that need to be calibrated concern the PACS event-points present in the NetLogo world. These points are colored with different shades of magenta (see Fig.S5), one for each of the $10$ PACS classes, and randomly move around the world during a simulation run with a frequency much greater than the simulation time step, that in our model corresponds to 1 year (in particular, each point shifts of 2 patches towards a random direction 73 times during each time step $t$ - i.e. with a frequency equivalent to 5 days).   

As explained in the MP, in our model the PACS event-points represent opportunities, ideas, encounters, intuitions, serendipity events, etc., which can periodically, and randomly, occur to a given researcher along her career. The relative abundance of points belonging to each PACS class is fixed in agreement with the information of the APS data set and it can be appreciated in the histogram shown in the left panel of Fig.\ref{world-zoom} (for example, it appears that the PACS code 70 is the most expressed, while the PACS code 90 is the less present). The total number $N_E$ of these points is one of the global parameters that have to be calibrated.  

The dynamical rules of the model, presented in detail in the MP, assume that the $N$ researchers, during their careers, are exposed to events and ideas which could trigger research lines, with the consequent articles production, along one or more different disciplinary fields according with the PACS numbers associated to each of the $N_E$ event-points. A given researcher $A_i$, depending on her interdisciplinary index $I^{APS}_i$, is sensitive only to the points corresponding to the numbers present in her PACS array $\vec{D}^{APS}_i$; let us define these points as 'special' for that researcher. Every year $t$, a check is performed over all the researchers in order to verify what and how many event-points would fall inside their "sensitivity circles", which represent the extension of their sensitivity to the special points and therefore influence the publication dynamics. In the right panel of Fig.\ref{world-zoom} is shown a zoom of the world, where three researchers, belonging to the three interdisciplinarity groups $L^{APS}_1$, $L^{APS}_2$ and $L^{APS}_3$, are reported together with their "sensitivity circles". The sizes of these circles are other three parameters that have to be calibrated through the comparison with real data.

\begin{figure}[t]
\begin{center}
\includegraphics[width=2.0 in,angle=0]{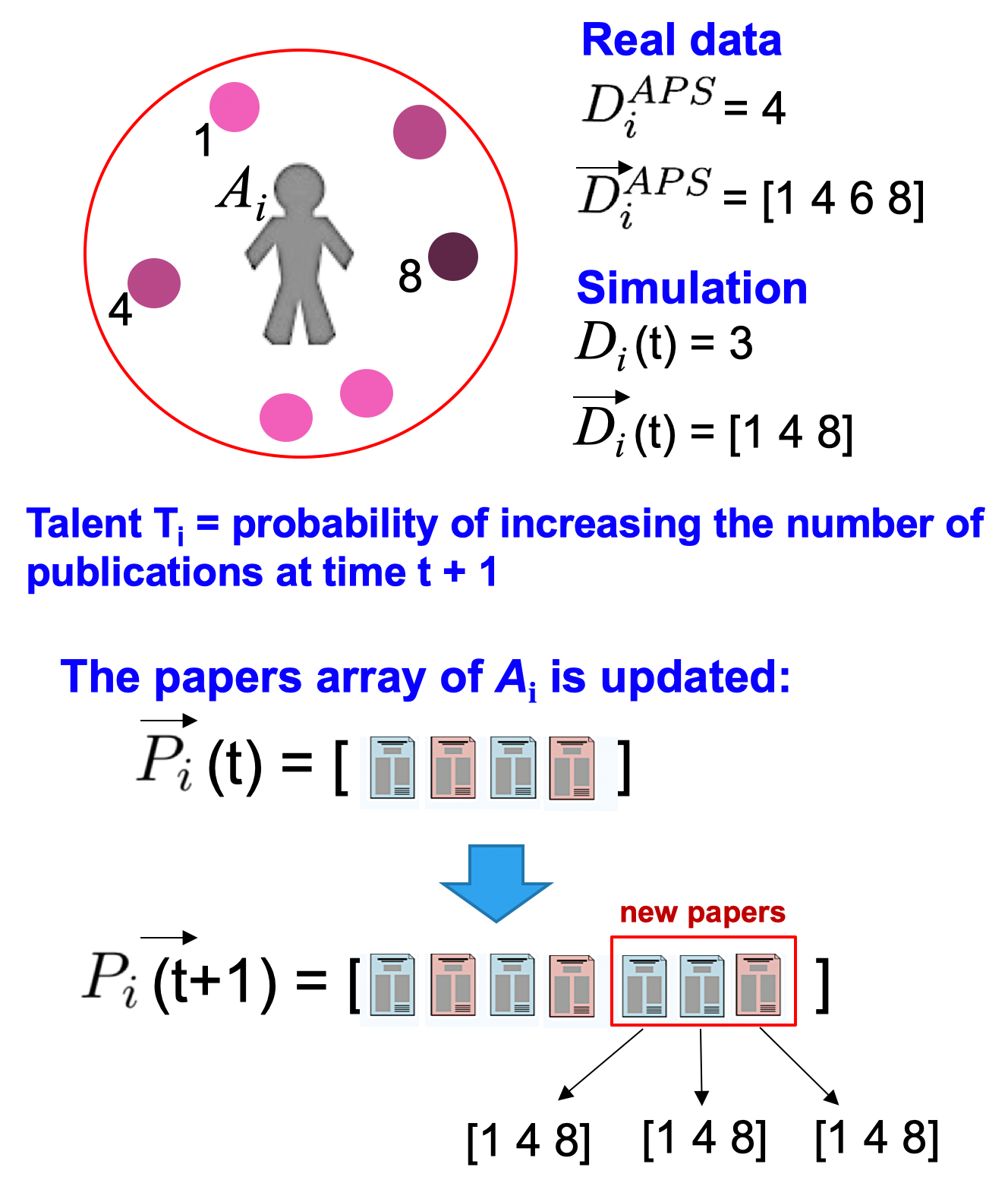}
\includegraphics[width=4.4 in,angle=0]{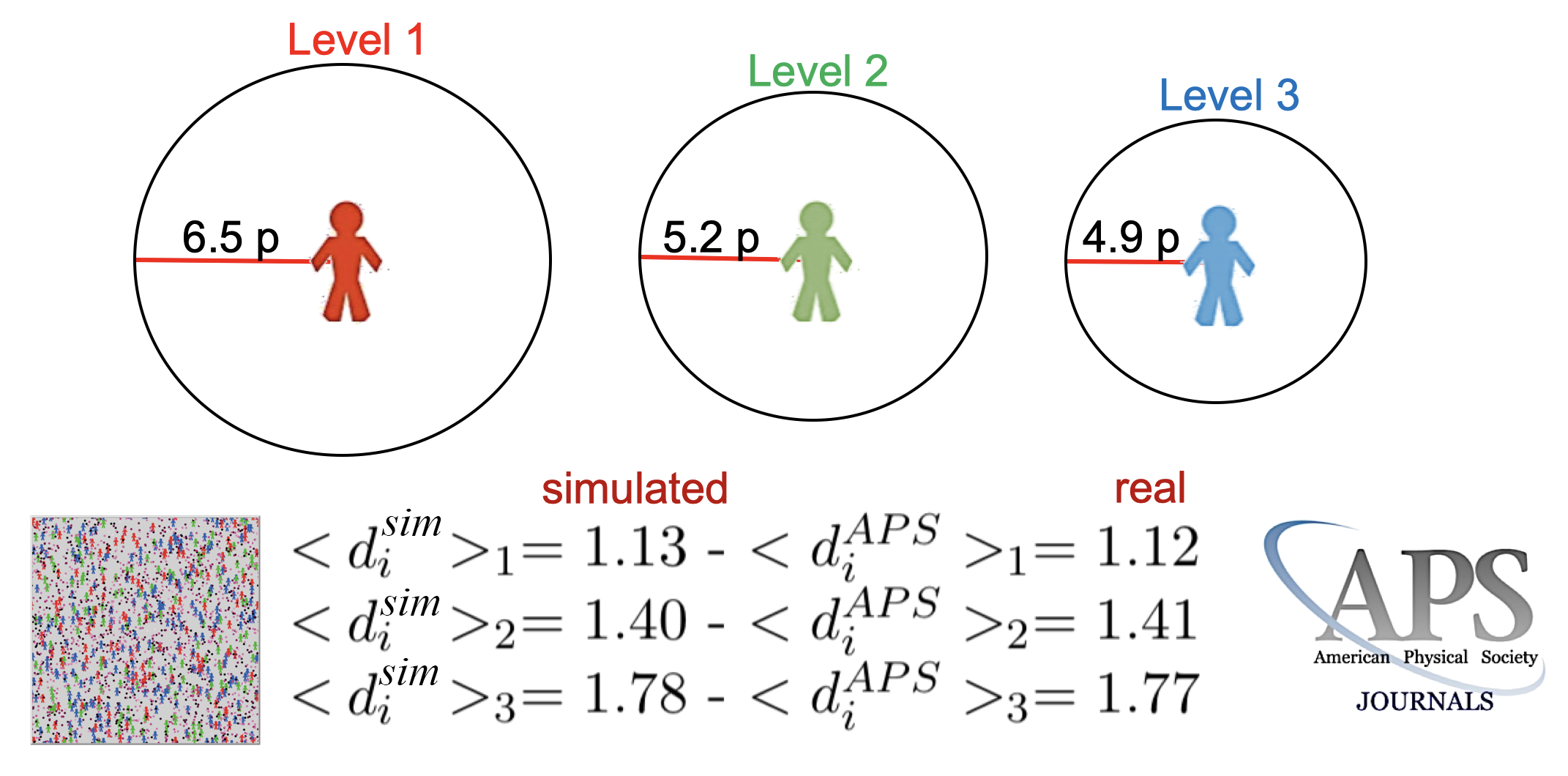}
\caption{\small 
(Left panel) An example of the dynamical rules for the publication of papers, see text. (Right panel) The total number $N_E$ of event-points and the radius of the sensitivity circles for the three groups of authors can be chosen by looking at the agreement between the average values of the simulated $d_i(t_{max})$ and the corresponding $d^{APS}_i$ obtained from the APS data set, see text.}
\end{center}
\end{figure}

In the left panel of Fig.17 we show an example of the publication dynamics for the generic author $A_i$. Let us suppose that $0 < D_i (t) \le D^{APS}_i$ is the number of special PACS points randomly falling in the sensitivity circle of $A_i$ at time $t$. In this example $D^{APS}_i=4$ but $D_i(t)=3$ since, among the four PACS numbers (1, 4, 6, 8) present in the array $\vec{D}^{APS}_i$ (real data), only three (1, 4, 8) do fall inside the circle. We can therefore define a temporary array $\vec{D}_i(t)$ containing these numbers. At this point, as explained in the MP, the considered researcher compares its talent $T_i$ with a random real number $r \in [0,1]$. Let us suppose that $r < T_i$: in this case the number $P_i(t)$ of her published papers increases of an integer quantity $\Delta P_i(t)$ randomly extracted from a Normal distribution with mean $m_{P_i}(t-1)  = \mu P_i (t-1)$ and standard deviation $\sigma_{P_i}(t-1) = \gamma P_i (t-1)$. The factors $\mu$ and $\gamma$ are other two global parameters (both $\le1$) that have to be determined by the comparison with real data (notice that these parameters are fixed in time and are common to all the authors, while $m_{P_i}(t-1)$ and $\sigma_{P_i}(t-1)$ are different for each author $A_i$ and are also variable in time, since they do depend on her past production at time $t-1$). Finally, all the new $\Delta P_i$ publications will be characterized by the PACS numbers contained in the array $\vec{D}_i(t)$. In the example of Fig.17 $\Delta P_i=3$, thus three new papers will be added to the papers array $\vec{P_i}(t-1)$ obtaining the new updated array $\vec{P_i}(t)$ where each of the new papers is characterized by the same three PACS numbers (1, 4, 8) -- in practice, for each paper a copy of the array $\vec{D}_i(t)$ is saved. 

The rationale behind these rules is twofold. On one hand, each researcher $A_i$ exploits the opportunities offered by the event-points falling in her sensitivity circle with a probability proportional to her talent, i.e. more talented authors have a greater a-priori probability of publishing new papers. On the other hand, the periodic increment in the number of publications is a constant fraction of the already published papers, i.e. the greater is the number $P_i(t-1)$ of existing publications at time $(t-1)$, the higher is the number $\Delta P_i$ of new publications at time $t$ (a sort of Matthew effect). Of course several approximations with respect to the reality have been assumed here. In particular, we assign the same PACS numbers to all the new papers published by $A_i$ at time $t$ and we do not consider coauthoring in the papers publication (each paper has a single author). This latter approximation contributes to produce an excess of published papers at the end of a simulation, but this is not a problem since we are interested in reproducing only the stylized fact represented by the shape of the papers distribution.        

\begin{figure}[t]
\begin{center}
\includegraphics[width=4.3in,angle=0]{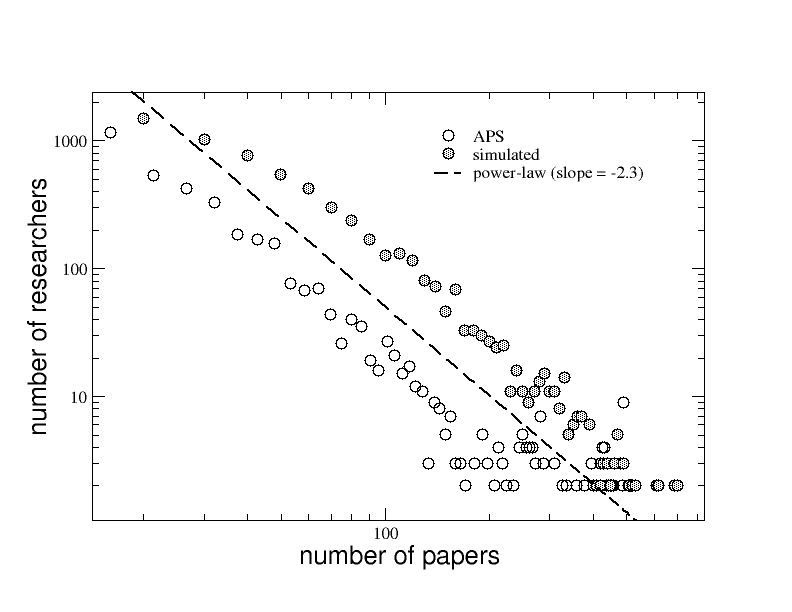}
\caption{\small 
Comparison between the papers distribution obtained from APS data set (open circles) and that obtained with the model simulation with $N_E=2000$, $\mu=1/5$ and $\gamma=1/4$ (full circles). The two distributions show a power-law behavior with the same exponent $-2.3$.  
}
\label{papers} 
\end{center}
\end{figure}

In order to choose the correct values for the global parameters previously introduced, i.e. the total number $N_E$ of event-points, the radius of the sensitivity circles and the factors $\mu$ and $\gamma$, we have run several simulation tests with different combinations of these parameters and compared the numerical results with the real APS data. 

First, we considered the averages $<d^{sim}_i>_g$, calculated over all the authors of the three groups ($g=1,2,3$), of the average number $d^{sim}_i(t_{max})$ of different PACS simultaneously present in their publications at the end of the simulation (i.e. at $t=t_{max}$) and compared them with the analogous real values $<d^{APS}_i>_g$ ($g=1,2,3$). It turned out that the values of $<d^{sim}_i>_g$ strictly depend on both the total number $N_E$ of event-points and the radius of the sensitivity circles. The choice of $N_E=2000$ and of a radius of $6.5$, $5.2$ and $4.9$ patches for the groups $L_1$,$L_2$ and $L_3$ respectively, was able to produce the best agreement with the APS data, with an error of $1\%$, as shown in the right panel of Fig.17. The decreasing size of the radius of the sensitivity circles for increasing interdisciplinarity levels, can be also justified by the evidence that the probability for a given researcher $A_i$ to find special event-points inside her sensitivity circle increases with $D^{APS}_i$, and therefore with the interdisciplinarity index $I^{APS}_i$, thus if we adopted the same size of the circles for the three groups $L_1$, $L_2$ and $L_3$, we would introduce a bias in favor of authors with medium and in particular with high interdisciplinarity level.     

Second, we were able to choose the correct values for the factors $\mu$ and $\gamma$ by comparing the simulated distribution of all the published papers (without distinctions among the interdisciplinarity levels) with the real one extracted from the APS data set. It turned out that the choice $\mu=1/5$ and $\gamma=1/4$ was able to produce a simulated papers distribution with a power-law behavior with the same slope (-2.3) of the real one (see Fig.\ref{papers}). Notice that, due to the constraints imposed by the calibration, these first results are very robust and do not depend on the details of the initial conditions of the simulations (i.e. do not depend neither on the particular realization of the distribution of talent among the agents, nor on the initial random position of both the agents and the event-points).

\begin{figure}
\begin{center}
\includegraphics[width=2.8in,angle=0]{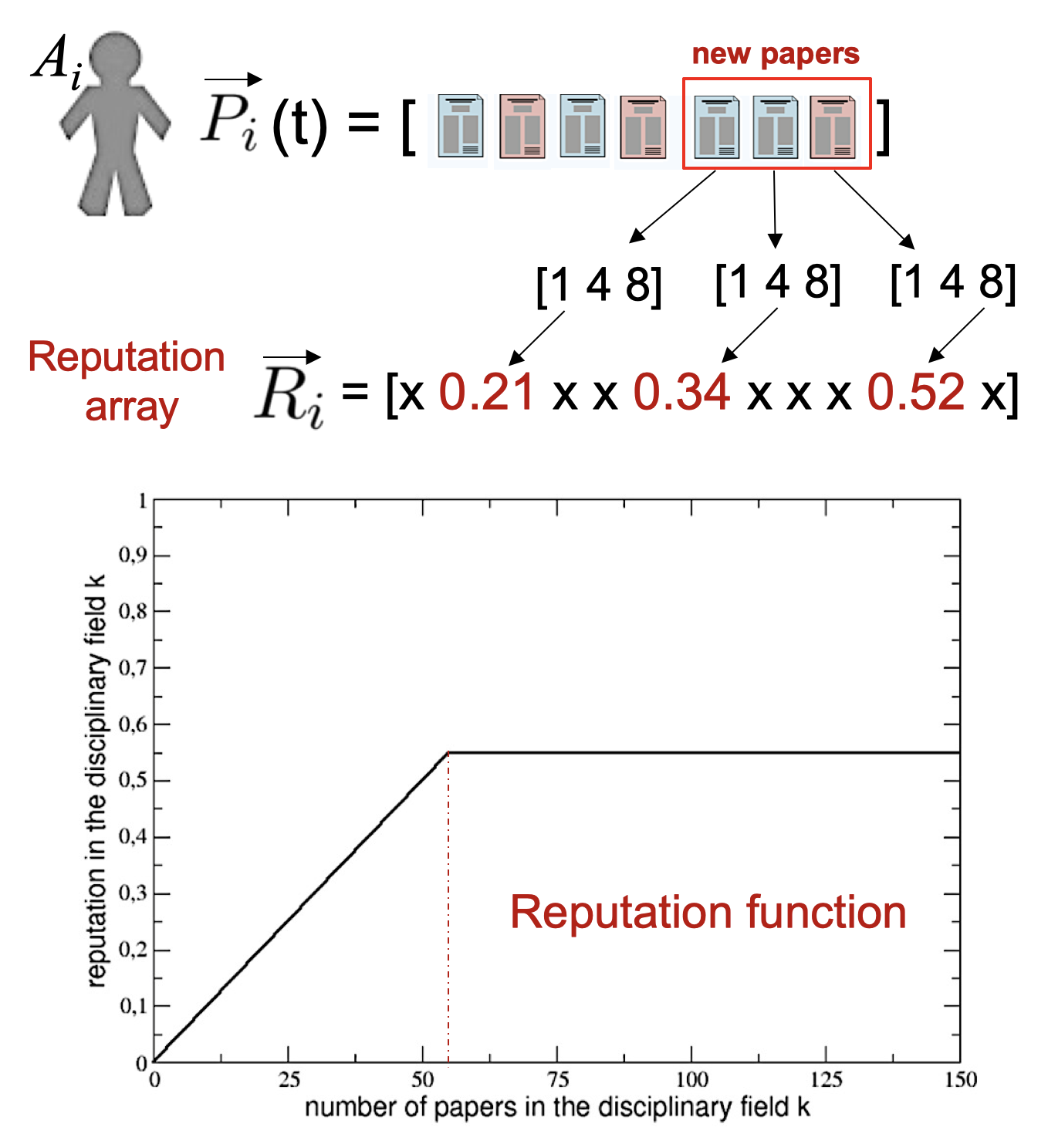}
\includegraphics[width=4.2in,angle=0]{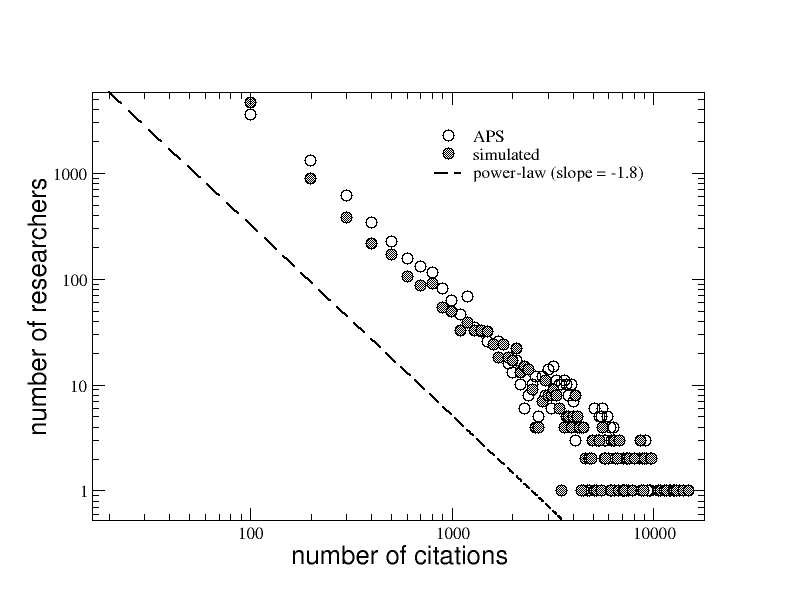}
\caption{\small 
(Left Panel) An example of the dynamical rules regulating the increase of reputation of an author $A_i$ in the fields corresponding to the PACS present in each new publication, see text. (Right Panel) Comparison between the citations distribution obtained from APS data set (open circles) and that obtained with the model simulation with $k=0.01$ and $y_{max}=0.55$ (full circles). The two distributions show a power-law behavior with the same exponent $-1.8$.  
}
\label{citations-simul} 
\end{center}
\end{figure}

Let us finally address the calibration of the citation dynamics for our model. As we have just seen, every year $t$ all the researchers have the chance to increase their number of publications. In correspondence of each new paper, author $A_i$ also increases her own reputation in each of the disciplinary fields indicated by the PACS numbers associated to that paper. As explained in the MP, each one of the 10 elements of the reputation array $\vec{R}_i(t)$ is a real number, included in the interval $[0,1]$, representing the reputation level reached by the researcher $A_i$ in the corresponding disciplinary field at time $t$ (see the top-left panel of Fig.S10 for an example). 

A plausible approximation to account the behavior of the reputation level $y$ of a generic author in a given field at time $t$ can be that of considering it as a semi-linear function of the number $x$ of papers published in that field at time $t$. In other words, we assume that $y$ does vary with $x$ following the function  

   \begin{equation*}
~
  y =    
  \begin{cases}
 k \cdot x ~~~~~~~~~~~~~   for ~~ x < x_{th}
  \\ y_{max} ~~~~~~~~~~~~   for ~~ x \ge x_{th}
   \end{cases}
   ~~~~
   \end{equation*}
              
where $k$ and $y_{max}$ are global parameters that, again, have to be calibrated with the real data, while $x_{th}$ is the abscissa of the inflection point (that depends on $y_{max}$).

Since, following the publication/citation dynamical rules explained in the MP, the total number of citations $C_i(t+1)$ reached by the author $A_i$ at time $t+1$ does depend on both her citation score $C_i(t)$ and her reputation array $\vec{R}_i(t)$ at time $t$ (Matthew effect), the choice of $k$ and $y_{max}$ does influence the citations distribution obtained at the end of a simulation (i.e. at $t=t_{max}$). Through several simulation tests, where different combinations of these parameters were adopted, we found that the values $k=0.01$ and $y_{max}=0.55$ (see bottom-left panel of Fig.S10) were able to produce a simulated overall citations distribution (without distinctions among the interdisciplinarity levels) that overlaps the analogous distribution obtained from the APS data set, following a power-law behavior with the same slope ($-1.8$, see the right panel of Fig.\ref{citations-simul}). Again, the constraints imposed by the comparison with the real data make these simulation results very robust, substantially independent from the initial conditions.

In conclusion, as last point to address, we also notice that - as observed in the MP - the calibrated agreement between the simulated averages $<d^{sim}_i>_g$ and the analogous real ones $<d^{APS}_i>_g$ for the three interdisciplinarity groups ($g=1,2,3$) do not ensure, of course, the correspondence of the individual $d^{sim}_i(t_{max})$ ($i=1,...,N$) of each agent-author at the end of a simulation with her initially assigned $d^{APS}_i$. Being the $D^{APS}_i$ fixed for all the authors during the simulation, this also implies that their initial value of the (real) individual interdisciplinarity index $I^{APS}_i$ can be different with respect to the corresponding one $I^{sim}_i(t_{max})=D^{APS}_i\times d^{sim}_i(t_{max})$ obtained at the end of the simulation. As a consequence, after a given simulation run, all the authors have to be reassigned -- on the basis of the same rules described in paragraph 1.1 -- to the three interdisciplinarity groups before calculating the corresponding papers and citations distributions (as those showed in the MP). We call these new groups $L_1^{sim}$, $L_2^{sim}$ and $L_3^{sim}$. It results that the number of authors belonging to $L_1^{sim}$, $L_2^{sim}$ and $L_3^{sim}$ is not exactly the same of the number of authors belonging to the original groups $L_1^{APS}$, $L_2^{APS}$ and $L_3^{APS}$, but typically the differences between the old and the new groups do not exceed $10\%$. In the simulation results presented in the MP, the sizes of the three new groups were, respectively, $N_1=2591$, $N_2=2383$ and $N_3=2329$. With respect to the original sizes shown in Table III, we notice that $L_1^{sim}$ slightly increased the number of its members, group $L_2^{sim}$ slightly decreased it, while group $L_3^{sim}$ leaved it relatively unchanged.

\end{document}